\documentclass[
 aps, pra,
 amsmath,amssymb,
 notitlepage,
 twocolumn,
superscriptaddress,
nofootinbib ]
{revtex4-2}

\usepackage{graphicx}
\usepackage{dcolumn}
\usepackage{bm}
\usepackage{braket}
\usepackage{amsthm}

\usepackage{xcolor}
\usepackage{todonotes}

\begin{document}

\title{A Localized Reality Appears To Underpin Quantum Circuits}

\author{Ken Wharton}
\affiliation{Department of Physics and Astronomy, San Jos\'e State University, San Jos\'e, CA 95192-0106}
\author{Roderick Sutherland}
\affiliation{Centre for Time, University of Sydney, NSW 2006 Australia}
\author{Titus Amza}
\affiliation{Department of Physics and Astronomy, San Jos\'e State University, San Jos\'e, CA 95192-0106}
\author{Raylor Liu}
\affiliation{Department of Physics and Astronomy, San Jos\'e State University, San Jos\'e, CA 95192-0106}
\affiliation{Department of Physics, University of Oregon, Eugene, OR 97404}
\author{James Saslow}
\affiliation{Department of Physics and Astronomy, San Jos\'e State University, San Jos\'e, CA 95192-0106}

\date{\today}

\begin{abstract}
Although entangled state vectors cannot be described in terms of classically realistic variables, localized in space and time, any given entanglement \textit{experiment} can be built from basic quantum circuit components with well-defined locations.  By analyzing the (local) weak values for any given run of a quantum circuit, we present evidence for a localized account of any circuit's behavior.  Specifically, even if the state is massively entangled, the weak values are found to evolve only when they pass through a local circuit element. They otherwise remain constant and do not evolve when other qubits pass through their circuit elements.  A further surprise is found when two qubits are brought together in an exchange interaction, as their weak values then evolve according to a simple classical equation.  The weak values are subject to both past and future constraints, so they can only be determined by considering the entire circuit ``all-at-once'', as in action principles.  In the context of a few basic quantum gates, we show how an all-at-once model of a complete circuit could generate weak values without using state vectors as an intermediate step.  Since these gates comprise a universal quantum gate set, this lends support to the claim that any quantum circuit can plausibly be underpinned by localized variables, providing a realistic, lower-level account of generic quantum systems.
\end{abstract}

\maketitle

\section{Introduction}

Unlike general relativity, quantum theory does not seem to describe a single, spacetime-localized account of what might actually be happening between measurements.  For instance, while any $N$-qubit state vector can be physically generated from $N$ interacting single-qubit worldlines in spacetime, the state vector itself is represented by $2^N$ complex numbers.  It might therefore seem that any spacetime-localized account of this system would require associating $2^N/N$ complex numbers with each qubit's worldline.  This is strikingly unrealistic, as each qubit's local description would have to grow exponentially with $N$, simply because of the existence of other physically-distant qubits.

But a mathematical description with exponential scaling need not correspond to an exponentially-complicated underlying reality.  After all, in classical statistical mechanics,  $N$-particle states also reside in an exponentially large configuration space.  In that case, the resolution is that the states represent incomplete knowledge, underpinned by $N$ localized particles in spacetime.  Finding a similar account, localized in spacetime, for $N$-particle entangled states has been identified as a grand challenge \cite{wharton2020}, one that would provide a novel way to reconcile quantum theory with spacetime-based general relativity.\footnote{This reconciliation would be distinct from standard quantum gravity approaches, as it would not treat quantum states as fundamental.}  While Bell’s Theorem has placed severe constraints on such a potential reformulation, this does not rule out ``future-input-dependent'' models \cite{wharton2020} where the actual history depends, in part, on the future measurement basis.  Any model along these lines would need to solve for entire histories ``all-at-once’’, drawing further comparisons with general relativity. 

The well-known quantum circuit framework (see Figure 1 for an example) is ideal for exploring this possibility.  Quantum circuits are modular; every possible quantum circuit can be built from a few basic one- and two-qubit gates, and the behavior of these simple gates is easy to analyze.  An obvious spacetime structure is also evident from the circuit itself.  The $N$ qubit ``wires’’ (the lines in Figure 1) can represent actual worldlines (say, paths of physically localized particles).  The two-qubit gates indicate where interactions occur: where the worldlines are physically brought together.  Furthermore, any complete quantum circuit has a pre-chosen measurement basis at its end, and this information is therefore available when implementing any analysis of the full circuit.  Even if the number of \textit{possible} measurement bases grows exponentially with $N$, in any complete circuit the number of \textit{actual} measurement bases is exactly one.  Thinking of the state vector as a look-up table with probabilities for all possible outcomes for all possible measurement bases, this table would grow exponentially with $N$.  But for the one future measurement that will actually happen, the underlying state of reality merely needs to scale linearly with $N$.  The analysis of this paper indicates that it is plausible that such a localized model, linear in $N$, underpins any quantum circuit. 

\begin{figure}[htbp]
\begin{center}
\includegraphics[width=8cm]{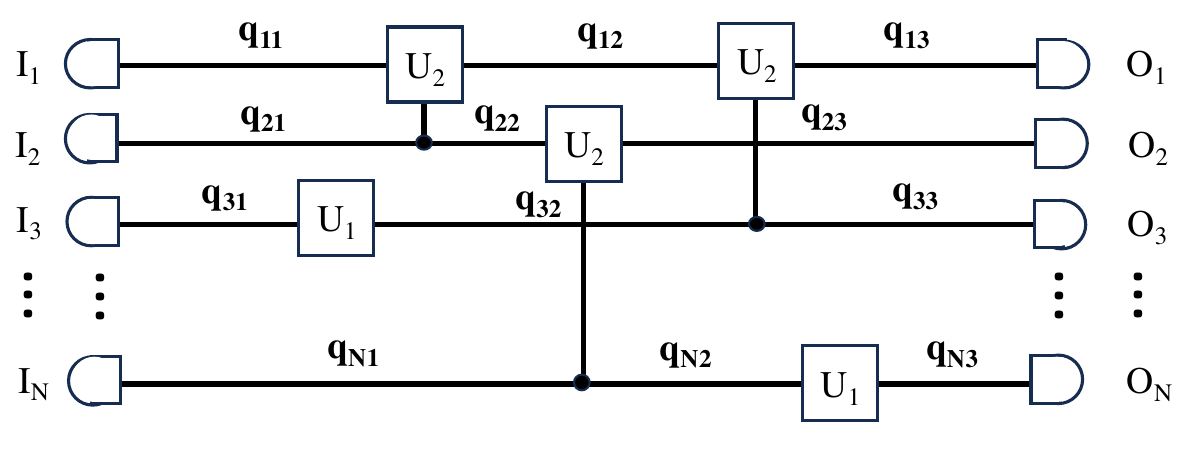}
\caption{An example of an N-qubit quantum circuit, with one- and two-qubit unitary gates ($U_1$ and $U_2$ respectively), classical inputs $I_i$ and classical outputs $O_i$.  The preparation and measurement gates are shown explicitly, indicating a complete circuit, and distinguishing the qubits from the classical inputs and outputs.  The question explored in this paper is whether an account of a complete circuit's behavior can be explained via localized variables $q_{ij}$ associated with the $j$th discrete segment of the $i$th wire, as shown.}
\label{default}
\end{center}
\end{figure}

The primary analysis of these circuits will be done using a well-known technique to extract the local properties of a quantum system: post-selected weak values.  These values have been independently derived in a number of different contexts \cite{roberts1978, AAV, sutherland1998}, and they can also be measured in the laboratory (by averaging over many identical runs of an experiment) \cite{exp1,exp2,exp3,exp4}.  Post-selected weak values naturally incorporate information about the eventual measurement basis, making them future-input-dependent.  It is not common to use local weak values to analyze properties of entangled states (see Section 7 of \cite{sutherland2022} for an exception), but they are naturally suited to do precisely that.  Most of the central results in this paper will come from analyzing the behavior of the weak values as they pass through various quantum gates, and noting that they behave just like local properties would be expected to behave.  For instance, in the midst of a continuous two-qubit gate, one can see that these weak values evolve according to a simple, local, differential equation, even if the quantum state in question is completely entangled.

Finally, we will present an example of a hidden variable model motivated by these results, obeying the same differential equations but also subject to certain global constraints.  Because the constraints depend in part on the future measurement basis, the model is future-input-dependent and is therefore not constrained by Bell/CHSH inequalities \cite{wharton2020}.  

Taken together, this paper presents several strands of evidence that quantum circuits act as if they have a local hidden structure.  The evidence paints an intriguing picture of what might become a full spacetime-localized reformulation of quantum theory.  As the number of qubits $N$ tends towards macroscopic systems, this hidden structure would merely increase linearly with $N$, not exponentially.  We argue that this economy of scale far outweighs the fact that the evidence points to a somewhat more complicated structure at the one- and two-qubit level.  But even without this argument, the evidence that local weak values obey simple classical equations can speak for itself.

\section{Post-selected Weak Values}

\subsection{Overview and Definitions}

We will consider a general quantum circuit that begins with a preparation procedure at time $t_i$, determining the initial state vector $\ket{i}$. The known circuit elements allow one to calculate the state vector at the time of final measurement $t_f$ as ${U}[t_f-t_i]\ket{i}$, where ${U}[t_f-t_i]$ is notation used to summarize the unitary operations of the circuit from time $t_i$ to time $t_f$.  Of course, this calculated state vector is usually not what is actually observed at time $t_f$.  Instead, some eigenstate $\ket{f}$ of the final measurement basis is inferred from the measurement outcome.  In standard QM, this is the newly collapsed state.  For any given run of any given quantum circuit, $\ket{f}$ provides additional information as to what has actually happened between times $t_i$ and $t_f$.

If this previous claim seems surprising, consider that in any classical situation with some unknown details, learning about outcomes will generally provide information about events prior to those outcomes.  In a quantum context, this fact has been formally proven -- both theoretically \cite{AAV, dressel2015} and in the laboratory \cite{exp1,exp2,exp3,exp4} -- in the context of post-selected weak values.  Given the input, output, and structure of a quantum circuit, a weak value $W$ can be defined for any Hermitian operator ${A}$ at any intermediate time $t$ according to
\begin{equation}
\label{eq:weak}
    W[A](t) =\frac{\bra{f}{U}[t_f-t]\,{A}\,{U}[t-t_i]\ket{i}}{\braket{f|{U}[t_f-t_i]|i}}.
    \end{equation}
Experimental validations of this expression involve actual ``weak measurements'' of ${A}$, which must be repeated many times to bring the signal out of the noise, and then ``post-selected'' for the particular outcome $\ket{f}$ of interest.  When this procedure is performed, the averaged post-selected results match the predicted weak value in the limit of minimal coupling between the system and the weak measurement device.
    
It must be noted that the above expression is mathematically well-defined even if no ``weak measurement'' is actually carried out.  As Dressel puts it \cite{dressel2015}, ``one then interprets Eq. (\ref{eq:weak}) as the best estimate of the (unmeasured) average value of ${A}$ \dots given only $\ket{i}$,$\bra{f}$, and [${U}$]''.  This means there is no restriction on simultaneously considering the weak values of non-commuting observables. For a single qubit, the weak values for all three Pauli operators $\sigma_x$, $\sigma_y$ and $\sigma_z$ can be calculated together. Inserting these three operators into Eq. (1) yields the x, y and z components of a weak value vector:
\begin{equation}
\label{eq:sdef}
   \bm{w}=(W[\sigma_x],W[\sigma_y],W_[\sigma_z]). 
\end{equation}
This vector $\bm{w}$ will be a focus of the analysis below.

In general, the weak values are complex numbers, so $\bm{w}$ can be decomposed into two 3-vectors, $\bm{w}=Re(\bm{w})+i Im(\bm{w})$, or else viewed as a single \textit{complex} 3-vector.  Weak values are not dependent on the global phase; any global phase adjustment to either $\ket{i}$ or $\ket{f}$ will appear in both the numerator and denominator of Eq. (\ref{eq:weak}) and will cancel.  The complex values of $\bm{w}$ are therefore not phase-dependent; there is an objective distinction between the vectors $Re(\bm{w})$ and $Im(\bm{w})$.

Note that the weak values are not restricted to the usual eigenvalues of the corresponding operator.  The values of $Re(W[A])$ can be much larger than the maximum eigenvalue of $A$ in cases where the probability of the actual outcome is small. Nevertheless, in the usual situation where the result of the final measurement is unknown and a weighted average is taken over the possible outcomes, the weak value $Re(W[A])(t)$ can then be shown to be exactly equal to the usual expectation value $\langle A\rangle (t)$. Hence the weak value tells us nothing new until we post-select a particular final outcome and look at the weak value for just that subset.

It is also possible to generate weak values associated with multiple qubits at once, because the operator $A$ in Eq. (\ref{eq:weak}) is not necessarily restricted to single-qubit operators.  But any such extension would defeat the purpose of this analysis, which is to find a realistic description of each individual qubit in a quantum circuit.  For this reason the weak values of such multi-qubit operators will not be considered in this paper; any use of the term ``weak values'' will be assumed to mean \textit{local} weak values, only concerning the properties of single qubits.

\subsection{Single Qubit Example}

Consider a qubit prepared in the state $\ket{i}=\ket{0}$ and then measured in the basis
\begin{eqnarray}
    \ket{f_+}&=&\cos \frac{\theta_0}{2} \ket{0} + \sin \frac{\theta_0}{2} \ket{1}\\
    \ket{f_-}&=&\sin \frac{\theta_0}{2} \ket{0} - \cos \frac{\theta_0}{2} \ket{1}
\end{eqnarray}
for some chosen value of $\theta_0$. 

Throughout this paper we will want to refer to such states as unit vectors on the Bloch sphere\footnote{One way to find these vectors from their corresponding states is $\ket{f}\bra{f}=(I+\hat{\bm{f}}\cdot\bm{\sigma})/2$, where $\bm{\sigma}$ is the usual vector of Pauli matrices.}, in this case $\hat{\bm{f}}_+$ and $\hat{\bm{f}}_-$.  Here the ``hat'' notation will be used for unit vectors corresponding to single-qubit states, so that $\hat{\bm{i}}$ represents the Bloch sphere representation of $\ket{i}$, etc.

With this notation, the above example corresponds to preparing a state in the $\hat{\bm{i}}=\hat{\bm{z}}$ direction and then measuring it to be in either the $\hat{\bm{f}}_+=\sin\theta_0\hat{\bm{x}}+\cos\theta_0\hat{\bm{z}}$ direction for the outcome $\ket{f_+}$, or in the opposite direction, $\hat{\bm{f}}_-=-\hat{\bm{f}}_+$ for the outcome $\ket{f_-}$.  The Hamiltonian in this example is zero; there is no time evolution of the qubit between preparation and measurement.  It follows from Eq. (\ref{eq:weak}) that $\bm{w}$ will therefore be constant as well.

 The weak values of the qubit at every point between preparation and measurement can be easily calculated for either measured outcome, $\ket{f_\pm}$.  Referring back to the vector defined in Eq. (\ref{eq:sdef}), each component can be calculated from (\ref{eq:weak}) by setting $A$ to the corresponding Pauli matrix.  For the two possible outcomes $\bra{f}=\bra{f_\pm}$, the two possible forms for $\bm{w}$ are: 
\begin{eqnarray}
    \bm{w}_+ &=& \begin{pmatrix} \tan(\theta_0/2) , & i\tan(\theta_0/2), &\hspace{5pt} 1 \hspace{10pt} \end{pmatrix} \\
    \bm{w}_- &=& \begin{pmatrix} -\cot(\theta_0/2) , & -i\cot(\theta_0/2), &\hspace{5pt} 1  \end{pmatrix} 
\end{eqnarray}
Note that both are unitless. If the qubit were a spin-1/2 particle, the only difference between these vectors $\bm{w}$ and the weak value vector of the spin-1/2 operators $(W[{S}_x],W[{S}_y],W[{S}_z])$ is that the latter vector would have an extra factor of $\hbar/2$.

Like $\hat{\bm{f}}_+$ and $\hat{\bm{f}}_-$, the real part of each weak value vector also lies in the $x-z$ plane, as shown in Figure 2. The real component of $\bm{w}$ bisects the angle between the initial and final states, and its length can also be found geometrically, as shown by the dashed lines in Figure 2.  A possible interpretation of this geometry will be discussed below.

\begin{figure}[htbp]
\begin{center}
\includegraphics[width=8cm]{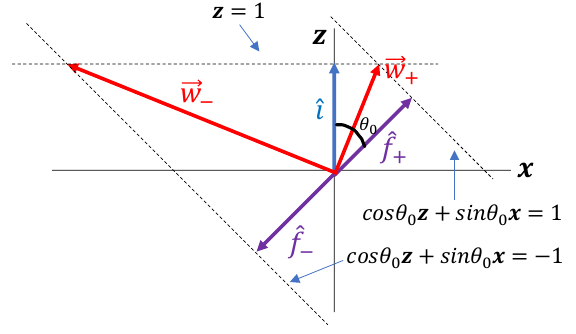}
\caption{The x-z plane of the Bloch sphere, showing the prepared state $\hat{\bm{i}}$, the two possible measured states $\hat{\bm{f}}_\pm$, and the real part of the two corresponding weak value vectors $\vec{\bm{w}}_\pm=Re(\bm{w}_\pm)$. These vectors lie at the intersection of two constraints corresponding to the initial preparation and final measurement (indicated by dashed lines).}
\label{default}
\end{center}
\end{figure}

The above example is actually quite general, as there is always a coordinate rotation which could transform any single-qubit prepare-and-measure scenario into this one.  As is well known \cite{AAV}, conventional normalization fails for weak values.  In this case, $Re(\bm{w}_+)^2=\sec^2(\theta_0/2)$ and $Re(\bm{w}_-)^2=\csc^2(\theta_0/2)$.  Curiously, these square magnitudes are inversely proportional to the probabilities of the corresponding outcomes.  Less surprisingly, the weighted average of $\bm{w}_+$ and $\bm{w}_-$ (using their Born-rule probabilities) is exactly $(0,0,1)$, with no imaginary part surviving.  This average matches $\hat{\bm{i}}$.

Even without conventional normalization, taking the imaginary portion of $\bm{w}$ into account, there is still a sort of ``hyperbolic normalization''. Treating the full $\bm{w}$ as a complex 3-vector, it is natural to define the complex scalar
\begin{eqnarray}
\label{sdots}
    \bm{w}\cdot\bm{w} &\equiv& [Re(\bm{w}) + iIm(\bm{w})]\cdot[Re(\bm{w}) + iIm(\bm{w})]\\
    &=& [Re(\bm{w})]^2 - [Im(\bm{w})]^2 + 2i Re(\bm{w})\cdot Im(\bm{w}).
\end{eqnarray}

Notice the minus sign on $[Im(\bm{w})]^2$; no complex conjugates have been taken.  Given this definition, one finds $\bm{w}\cdot\bm{w}=1$ (with a zero imaginary part) for both $\bm{w}_+$ and $\bm{w}_-$.

In this single-qubit example it is not hard to find further relationships between $\bm{w}$, the preparation vector $\hat{\bm{i}}$, and the measured vector $\hat{\bm{f}}$.  Certainly it is evident that $\bm{w}\cdot\hat{\bm{i}}=1$.  Less obviously, half-angle formulas reveal $\bm{w}\cdot\hat{\bm{f}}=1$, for either measured outcome -- this should also be evident from Fig. 2.  At the preparation, one always finds
\begin{equation}
\label{eq:sprep}
    \bm{w}=\hat{\bm{i}}+ i(\hat{\bm{i}}\times\bm{w}).
\end{equation}
At the final measurement, the cross product is reversed:
\begin{equation}
\label{eq:smeas}
     \bm{w}=\hat{\bm{f}} + i(\bm{w}\times\hat{\bm{f}}).
\end{equation}
Combined with $\bm{w}\cdot\bm{w}=1$, these last two equations uniquely determine the constant weak value vector $\bm{w}$ for any pair of sequential measurements on a single non-interacting qubit.

\subsection{Discussion}

Weak values have no widely accepted interpretation, but Roberts \cite{roberts1978} originally proposed that they could represent ``part of objective reality", some hidden feature of the actual quantum system.  This viewpoint has also been advocated and extended by Sutherland \cite{sutherland1998,sutherland2008,sutherland2017}.  

It is important to note a crucial distinction between these proposals and the so-called ``Two State Vector Formalism'' (TSVF) \cite{tsv}, in which the focus is on the (unlocalized) state vectors as elements of reality, as opposed to the (localized) weak values themselves.  Analysis of TSVF sometimes implies the reality of weak values, as something which the ``particles are allowed to possess'' \cite{aharonov2015}.  But more often TSVF treats weak values as arising at the time of weak measurements, rather than as continuously existing properties. In that view, weak values are then seen as a consequence of the state vectors rather than an explanatory resource in their own right.

If the goal laid out in the introduction is possible to achieve, then it should also be possible to make sense of the weak values on their own terms, without any reference to wavefunctions or state vectors at all.  This is essentially what was accomplished in the previous subsection (II.B) for a single non-interacting qubit.  True, the state vector viewpoint was first used to compute the weak value $\bm{w}$, but the resulting analysis yielded novel rules which made the state vector analysis superfluous.  With $\bm{w}$ remaining constant from preparation to measurement, it becomes possible to compute the intermediate $Re[\bm{w}]$ using only the initial preparation and the final measurement result, as shown geometrically in Figure 2.  Alternately, using Eqs. (\ref{eq:sprep}) and (\ref{eq:smeas}) along with $\bm{w}\cdot\bm{w}=1$, the complex weak value vector $\bm{w}$ can be determined without ever defining a state vector at all.

A physical interpretation of this weak value is also evident, at least for the case where the qubit is a spin-1/2 particle.  The initial preparation for this example can now be interpreted as choosing the result of an initial spin measurement in the $\hat{\bm{z}}$ direction, fixing the $z$-component of the weak value $\bm{w}\cdot\hat{\bm{z}}=1$.  No other component was measured at this moment, so there is no initial constraint on any other component of $\bm{w}$, or even its total magnitude.  The final measurement then constrains a \textit{different} component of $\bm{w}$ to be $\pm 1$ (either $\hat{\bm{f}}_+$ or $\hat{\bm{f}}_-$).  For each possible result, the smallest vector that conforms to both of these constraints, without changing between measurements, is precisely $Re(\bm{w}_\pm)$.  Such a hidden structure of a single spin-1/2 particle, as described by the weak value formalism, has previously been explored in Sec. 6 of \cite{sutherland2022}.  

While it is true that state vectors were used for calculating the probabilities of the two possible outcomes, even this use of the state vector can be easily discarded in this particular example.   As noted above, the Born rule could be replaced with a global postulate that assigns probabilities directly to the weak values:
\begin{equation}
\label{eq:prob1}
    P(\bm{w}) \propto \frac{1}{Re(\bm{w})^2}.
\end{equation}
Therefore, since one knows the two possible outcomes for any setting, and each outcome is associated with a computable $\bm{w}$, one could use this rule to determine the absolute probabilities of the two outcomes.  (This would not require normalizing the weak values, but rather simply using the fact that probabilities always sum to one.)

One objection at this point might be that such a weak-value-based account would not be an ``explanation'' in a traditional sense.  Indeed, if one demanded a dynamical explanation (starting from certain inputs in the past, solving dynamical equations, and then computing the outcome from that result), this approach might seem inadequate.

But for most \textit{physicists}, at least, it is not difficult to shift into an all-at-once mindset where explanations do not have to come in this form \cite{wharton2014,stuckey,adlam2022,chen2022}.  Action principles, in both classical and quantum theory, are analyzed ``all at once'' in exactly this manner, using both initial and final boundary constraints.  Much of general relativity also falls in this category.  If normal-mode cavity analysis and Ising-model logic can be applied ``all at once'' over large spatial regions, and we are thought to live in a universe without a strict distinction between temporal and spatial directions, then this same style of logic could certainly be applied to complete spacetime histories.  To reject such models as improper or non-explanatory would perhaps be an overly anthropocentric reaction, based more on our personal experience of time than on our theoretical understanding of spacetime.\cite{uinac}

Still, there is an important technical difference between the weak value model described above and these other examples.  Action principles use a final boundary constraint, but that constraint is not a controllable \textit{setting}.  For the above account to work, the weak values of the system would have to be correlated with the final setting choice $\theta_0$.  Different choices of the final measurement direction $\theta_0$ would select for different possible outcomes $\ket{f}$, and would therefore also select for different hidden states $\bm{w}$ at times stretching back to the initial preparation.  Such a situation has been termed ``future-input dependent'' as defined in the Introduction, and is evidently ``retrocausal'' by any reasonable definition of the term.\footnote{Mere correlation with an \textit{outcome} is not retrocausal in any way; the issue is that the weak value is here correlated with the future \textit{setting}.  Sometimes the term ``post-selection'' is confused with retrocausation, but unless one is actually imposing future boundary constraints, updating knowledge of the past based on a future outcome is completely standard in both classical physics and everyday reasoning.}  

But such retrocausality is not specific to a realistic interpretation of weak values.  Well-known results from Bell's Theorem independently motivate a correlation between the hidden variables and the future settings \cite{wharton2020}.  Any model which allows the future setting to constrain past hidden variables could potentially restore a localized and covariant explanation of how entangled particles maintain their Bell-inequality-violating correlations. This might roughly be viewed as an effect which zigzags through the past, but is almost certainly best analyzed all-at-once, as a history constrained by both past and future boundary constraints in a time-symmetric manner.  By using weak values to flesh out this general idea, it becomes possible to carefully analyze what this might actually look like at the hidden-variable level.

 We now turn to the Roberts/Sutherland proposal that the weak values may be the best clue we have as to what is really going on inside a quantum system.  The essential idea is to calculate the \textit{localized} weak values from a generic entangled state, and then attempt to make sense of them as realistic entities in their own right.  As argued in the Introduction, if it is possible to find a dynamically-local description of an $N$-qubit quantum circuit which scales linearly (rather than exponentially) with $N$, being forced to take the future measurement basis into account might be a relatively small price to pay.

\section{Individual Qubits in Quantum Circuits}

\subsection{Localized Weak Values}

Consider a generic N-qubit quantum circuit, where at every moment the full quantum state is generally entangled. Even for such an entangled state, a local weak value for each individual qubit can be defined as in Section II, so long as one uses an operator ${A}$ having the tensor product form ${A}=B\otimes I$.  Here $B$ is a single qubit operator (say, one of the Pauli matrices $\sigma_j$), and $I$ is the identity operator of the appropriate dimension to act on the remaining $N-1$ qubits.

The complex-valued vector of weak values $\bm{w}$ was defined in Eq. (\ref{eq:sdef}) for a single qubit.  When considering an N-qubit system, possibly in an entangled state, the analogous expression for a single qubit of interest (say, qubit ``a'') can be generalized to
\begin{equation}
\bm{w}_a=(W[\sigma_x\otimes I],W[\sigma_y\otimes I],W[\sigma_z\otimes I]).
\end{equation}

Such a procedure will work for any individual qubit; when more than one qubit is considered, the particular qubit will be noted with a subscript, such as $\bm{w}_a$, $\bm{w}_b$, etc.  Of course, there are also operators that refer to multiple qubits and so are not of the form $A=\sigma_j\otimes I$, but only weak values calculated with this special sort of operator are candidates for a localized, hidden description of what might be happening. 

The question at hand is whether the weak values evolve according to local conditions -- whether they are ``dynamically local''.  Any hope for such dynamic locality would be dashed if the single-qubit weak values calculated in this way depended upon the evolution of distant qubits in the same entangled state.  Equivalently, the worry might be that the weak values might reveal a special ``plane of simultaneity'', reflecting the reference frame in which the state vectors are defined, preventing any possible extension to a Lorentz covariant theory.  The next subsection will show that these potential concerns do not, in fact, occur.  

\subsection{Dynamic Locality for Entangled Qubits} 

In the single-qubit example of Section II.A, we saw that the weak value vector remained constant between preparation and measurement, so long as there was no Hamiltonian acting on it.  If there is a hidden local description underpinning the qubit of interest, $\bm{w}_a$ should also remain constant for any N-qubit Hamiltonian of the form $H_N=I\otimes H_{N-1}$, even if the other $N-1$ entangled qubits are evolving due to some Hamiltonian $H_{N-1}$.  In other words, given a Hamiltonian of this form, the weak value $W[\sigma\otimes I]$ should stay constant for any single-qubit operator $\sigma$.

To test this proposal, all one needs is to represent the unitary time-evolution operator in the form $U_N=I\otimes U_{N-1}$ and calculate the relevant weak values $W[\sigma\otimes I]$ at two different times.  Before the evolution, this yields
\begin{equation}
    W[\sigma\otimes I](t_i) =\frac{\bra{f}U_{N}(\sigma\otimes I)\ket{i}}{\braket{f|U_{N}|i}}.
\end{equation}
Meanwhile, at the final measurement time $t_f$, the same weak value is found via
\begin{equation}
    W[\sigma\otimes I](t_f) =\frac{\bra{f}(\sigma\otimes I)U_{N}\ket{i}}{\braket{f|U_{N}|i}}.
\end{equation}
Evidently\footnote{This follows because, if $A$ and $B$ are operators in the same Hilbert subspace (and the same is true for $C$ and $D$), then $[A\otimes C,B\otimes D]=[A,B]\otimes [C,D]$.}, $\sigma\otimes I$ commutes with $U_N=I\otimes U_{N-1}$.  This means that the above two expressions are identical, so the weak values will remain constant along any wire, even if the qubit is part of an entangled state. 

\subsection{Single-Qubit Gates}

Having the weak values remain constant on the wires is not enough.  They will also have to respond in a reasonable way when they encounter a single-qubit gate $U_1$, even with other qubits in the entangled state.  In this case, the relevant unitary time-evolution operator would be $U_1\otimes U_{N-1}$.  Picking out an arbitrary component of the weak value vector $\bm{w}_j$ for the qubit of interest (corresponding to some Pauli matrix $\sigma_j$), the weak value before the $U_1$ gate is equal to
\begin{equation}
    W[\sigma_j\otimes I](t_i) =\frac{\bra{f}(U_1\otimes U_{N-1})(\sigma_j\otimes I)\ket{i}}{\braket{f|(U_1\otimes U_{N-1})|i}}.
\end{equation}
This will no longer be the same as the same weak value after the gate. 

However, notice that after the gate there is a different weak value component, $W[\sigma_f\otimes I]$ which would be equal to the above expression.  This component could be calculated according to
\begin{equation}
    W[\sigma_f\otimes I](t_f) =\frac{\bra{f}(\sigma_f\otimes I)(U_1\otimes U_{N-1})\ket{i}}{\braket{f|(U_1\otimes U_{N-1})|i}}.
\end{equation}
A short calculation reveals that these two equations will be identical if $U_1 \sigma_j = \sigma_f U_1$, or, equivalently, if $\sigma_f=U_1\sigma_j U_1^{-1}$.

This means that the weak value $W[\sigma_j]$, originally associated with the $\sigma_j$ direction, has now been mapped to $W[\sigma_f]$ (associated with the $\sigma_f$ direction) after passing through the single-qubit gate. And the relationship between $\sigma_j$ and $\sigma_f$ is just a rotation, exactly what the gate $U_1$ is known to do to any Bloch-sphere vector. In other words, if one knew the weak value vector $\bm{w}$ going into a single qubit gate corresponding to a known rotation, the resulting weak values could be calculated simply by subjecting the weak value vector $\bm{w}$ to precisely that same rotation.  This rotation depends only on $U_1$, not on the $U_{N-1}$ evolution affecting the other qubits. Hence the weak value evolution through the gate is independent of what is simultaneously happening to the rest of the entangled state.


\subsection{Measurements}

In the single qubit case, we saw that the weak value vector $\bm{w}$ was partially fixed in the direction $\hat{\bm{n}}$ corresponding to the measured state.  Specifically, we found $\bm{w}\cdot\hat{\bm{n}}=1$, with no imaginary component in that direction.  Here we will briefly show why this is always the case, even in a multi-qubit circuit. 

Suppose the initial state $\ket{i}$ is subjected to any unitary evolution $U_N$, and then one qubit is measured.  The final measurement operator for the qubit of interest is $\sigma_f$, with eigenfunctions that solve $\sigma_f\ket{f_\pm}=\pm\ket{f_\pm}$.  A measurement of only that qubit (corresponding to the full operator $\sigma_f\otimes I$) will therefore result in the partially-separable final state $\ket{f}=\ket{f_\pm}\otimes\ket{\psi_{N-1}}$. 

We are interested in the weak value component for that one qubit, $\bm{w}\cdot\hat{\bm{f}}_+=W[\sigma_f\otimes I]$, just before measurement.  Since the evolution $U_N$ has already occurred, this yields
\begin{equation}
    W[\sigma_f\otimes I](t_f) =\frac{\bra{f}(\sigma_f\otimes I)U_N\ket{i}}{\braket{f|U_N|i}}.
\end{equation}

However, since $\bra{f}=\bra{f_\pm}\otimes\bra{\psi_{N-1}}$, and $\bra{f_\pm}\sigma_f=\pm\bra{f_\pm}$, this simplifies to 
\begin{equation}
    W[\sigma_f\otimes I](t_f) =\frac{\pm\bra{f}U_N\ket{i}}{\braket{f|U_N|i}} = \pm 1.
\end{equation}
Therefore, if $\ket{f_+}$ is measured, then $\bm{w}\cdot\hat{\bm{f}}_+=1$.  If $\ket{f_-}$ is measured, $\bm{w}\cdot\hat{\bm{f}}_+=-1$, but this is the same as $\bm{w}\cdot\hat{\bm{f}}_-=1$.  This matches the single-qubit results from the previous section.

The above analysis can also be easily inverted to apply at preparation, where (unlike at measurement) one has additional control to choose the eigenvalue.

\subsection{Discussion}

To summarize the above results, a qubit on any ``wire'' stretching between two gates in any quantum circuit will always have a constant weak value vector $\bm{w}$ (both imaginary and real parts) until it hits a gate.  This is true even if it is part of an entangled state, and even if the other qubits are simultaneously passing through gates of their own. 
 Furthermore, when one qubit passes through a single-qubit gate which implements a (Bloch sphere) rotation, the corresponding weak value will rotate in the same manner regardless of whether there is entanglement with other qubits or not. This is true for both real and imaginary parts.  These results are precisely the behavior one would expect if the weak values represented a hidden localized account of what was actually happening in any given quantum circuit.

It is important to emphasize that this analysis concerns the dynamical evolution of the weak values as they pass through a given quantum circuit, not counterfactual changes between different circuits.  Suppose circuits $C_1$ and $C_2$ start out in an identical manner, but they then use a different single-qubit gate at one point in mid-circuit.  In any all-at-once analysis of the two circuits, that difference at one point might very well lead to differences in the weak values at every point.  By comparing $C_1$ and $C_2$, basic causal reasoning then tells us that an intervention at one point (at one gate) can have effects on distant qubits, in apparent contradiction to the above analysis.

But there is no contradiction here.  The above proofs show that in both circuits $C_1$ and $C_2$ the weak values always stay constant on the circuit wires, and only dynamically change when they pass through a gate.  The apparent nonlocal influence described above (and discussed, for example, in \cite{braverman}) is not a dynamical change, but a counterfactual ``change'', only evident when comparing the complete runs of two different circuits.  Evidently, this sort of counterfactual influence must be allowable in any sort of model which can violate the Bell/CHSH inequalities and thereby agree with observed entanglement correlations.

\section{Dynamic Locality for Two-Qubit Gates}

\subsection{A Universal Two-Qubit Gate}

When two identical qubits interact via the exchange interaction, with a Hamiltonian of the form $H_{ex}=J({\sigma}_{x}\otimes{\sigma}_{x}+{\sigma}_{y}\otimes{\sigma}_{y}+{\sigma}_{z}\otimes{\sigma}_{z})\equiv J(\bm{\sigma}\otimes\bm{\sigma}$), they evolve in a way that is equivalent to passing through a two-qubit SWAP$^\alpha$ gate.  Here $J$ is a constant which determines the coupling energy.  In the $\ket{00},\ket{01},\ket{10},\ket{11}$ basis, to within a global phase, the corresponding unitary operator for this exchange interaction is
\begin{equation}
\label{eq:Uex}
U_{ex}[\alpha] = \begin{pmatrix} 1 & 0 & 0 & 0 \\
0 & \frac{1+e^{i\pi\alpha}}{2} & \frac{1-e^{i\pi\alpha}}{2} & 0 \\
0 & \frac{1-e^{i\pi\alpha}}{2} & \frac{1+e^{i\pi\alpha}}{2} & 0 \\
0& 0&0&1
\end{pmatrix}
\end{equation}
Here $\alpha$ is used to compute the net result of the total interaction, so it is therefore a product of the interaction strength $J$ and the total time over which the interaction persists.  In this section it will be useful to think of $J$ as being a constant over the duration of the interaction, with $\alpha$ varying to indicate this duration, scaled such that $\alpha=1$ corresponds to a full SWAP of the two qubits. 

Turning on the interaction for only half the SWAP time ($\alpha=0.5$) corresponds to a $\sqrt{\text{SWAP}}$ gate.  This gate can be used to entangle two separable qubits and two $\sqrt{\text{SWAP}}$s (combined with single qubit gates) can generate the more common CNOT gate.  It is well known that the CNOT and single-qubit gates comprise a universal set for generating any possible quantum circuit, so the same must be true for the $\sqrt{\text{SWAP}}$ gate when combined with the single-qubit gates from the previous section.

Apart from generating a universal set of gates for quantum computation, we focus on the SWAP$^\alpha$ gate for two additional reasons.  First, the exchange interaction is the most ``natural'' interaction for two charged spin-1/2 systems; such a Hamiltonian can be implemented simply by bringing two such particles to the same average location.  More importantly, when thinking of $\alpha$ as a time parameter, the SWAP$^\alpha$ represents a \textit{continuous} gate (as contrasted with a complete CNOT operation).  Since weak values can be computed without actually making a weak measurement, one can compute ``movies'' of the weak values as this SWAP$^\alpha$ interaction proceeds simply by changing the values of $\alpha$ in the gates before and after the point at which the weak values are calculated.  Carrying out this analysis will indicate that the weak values are evolving according to a simple classical equation, even as quantum theory says that the qubits are becoming entangled. 

\subsection{Results: Dynamic Locality}

To get a sense of the behavior of the weak values in an exchange interaction, we first turn to numerical modeling.  Our code chooses a random two-qubit state $\ket{i}$ for the preparation, then evolves these two qubits with a SWAP$^\alpha$ gate (for any chosen value of $\alpha$).  The gate output is measured in a manner that corresponds to another random two-qubit state $\ket{f}$.  At this stage, there is no restriction on $\ket{i}$ or $\ket{f}$; they can each be either entangled or separable.

To parameterize the evolution of the weak values inside the SWAP$^\alpha$ gate, it is convenient to use a unitless time parameter $\tau$, ranging from $\tau=0$ at the start of the gate to $\tau=\alpha$ at the end.  (Recall that we are keeping the strength of the exchange interaction constant, so $\tau$ evolves linearly with time.)  Using the above notation, the weak value vector $\bm{w}$ for qubits $a$ and $b$ at any point in the midst of the SWAP$^\alpha$ gate can be found according to 
\begin{eqnarray}
    \bm{w_a}(\tau)&=&\frac{\bra{f} U_{ex}[\alpha-\tau] (\bm{\sigma}\otimes I)U_{ex}[\tau]\ket{i}}{\braket{f|U_{ex}[\alpha]|\psi}}\\ 
    \bm{w_b}(\tau)&=&\frac{\bra{f} U_{ex}[\alpha-\tau] (I\otimes \bm{\sigma})U_{ex}[\tau]\ket{i}}{\braket{f|U_{ex}[\alpha]|\psi}}
\end{eqnarray}

As before, these are both complex 3-vectors; complex because they are weak values and 3-vectors because of the definition of the vector of Pauli matrices $\bm{\sigma}=(\sigma_x,\sigma_y,\sigma_z)$.  Also note that unlike the many-qubit identity operators in Section III, the $I$ here applies to only a single qubit.

Numerical simulation of these two complex vectors reveals that the average $\bm{w}_{avg}=(\bm{w_a}+\bm{w_b})/2$ always remains constant over time, for each of the six components (3 real and 3 imaginary).  More interestingly, each of the six components always oscillate around the corresponding component of $\bm{w}_{avg}$ as a simple harmonic oscillator.  Figure 3 shows a completely typical result for one of these components, at a series of intermediate points, for $0\le\alpha\le 2.3$.  The fit line is a sinusoidal oscillation around the average, and the results always lie on this line to within expected numerical error.

\begin{figure}[htbp]
\begin{center}
\includegraphics[width=8cm]{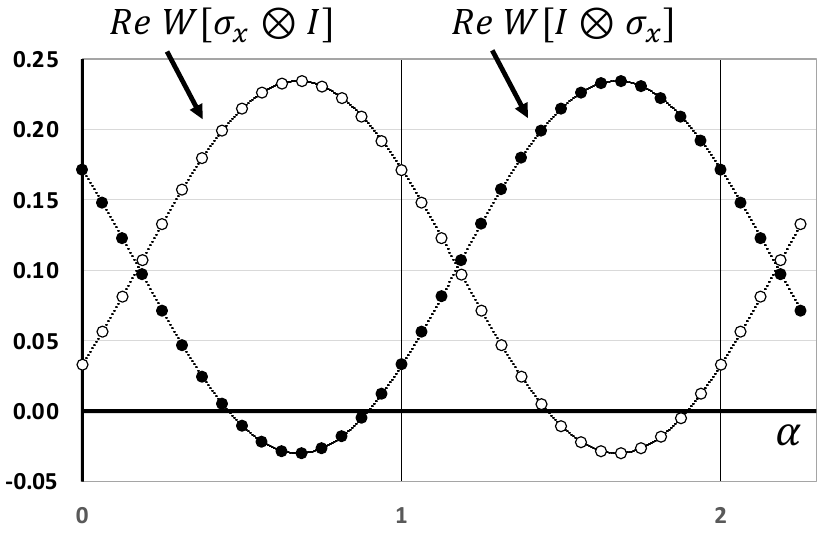}
\caption{This graph shows a typical pair of local weak value components for two qubits at various points during an exchange interaction. The essentially-perfect fit indicates that they are obeying a basic harmonic oscillator equation.  As required, the components always ``SWAP'' at $\alpha=1$, and return to their original state at $\alpha=2$.}
\label{default}
\end{center}
\end{figure}

We can infer from these results that during an exchange interaction, the weak values do indeed obey a differential equation of the precise form
\begin{eqnarray}
\label{eq:eoms}
    \frac{d^2\bm{w_a}}{d\tau^2} &=&   \frac{\pi^2}{2} (\bm{w_b}-\bm{w_a})\\
    \frac{d^2\bm{w_b}}{d\tau^2} &=&   \frac{\pi^2}{2} (\bm{w_a}-\bm{w_b}).\nonumber
\end{eqnarray}

However, merely knowing the initial values of $\bm{w_a}$ and $\bm{w_b}$ is not sufficient to solve these equations.  One would also need to know additional information, such as $d\bm{w_a}/d\tau$ and $d\bm{w_b}/d\tau$ at the beginning of the interaction.  Another path forward would be to use both the initial and final values of $\bm{w_a}$ and $\bm{w_b}$ to solve these equations, as is commonly done in action extremization problems.

With these numerical results in hand (\textit{i.e.}, knowing
what to look for), it becomes possible to find general solutions for the weak value components in the $SWAP^\alpha$ gate and show that they solve Eq. (\ref{eq:eoms}).  This was checked analytically by one of us (RS), thereby providing confirmation of the above equations.  A different analytical proof, starting with Eq. (\ref{eq:weak}), can be found in Appendix A. 

These results were derived for a two-qubit system, but they were also checked for a multi-qubit system, where the state was fully entangled but only two qubits were interacting.  The success of this extension follows from generalizing the results in Section III, parsing the system into a 2-qubit subspace and a $(N\!\!-2)$-qubit subspace, $U=U_{ex}\otimes U_{N-2}$.  As before, any entanglement is irrelevant to the dynamic locality observed in the local weak values.

\subsection {Other Results}

In the single-qubit case from Section II, we saw that the weak values always obeyed the relationship $\bm{w}\cdot\bm{w}=1$.  This fails in general for multi-qubit entangled states, but does continue to hold in at least two special cases.  The first case is when one or more qubits is not entangled with the others, according to the forward-evolved state vector $U[t-t_i]\ket{i}$.  If any single qubit is separable from the others, such that $U[t-t_i]\ket{i}=\ket{n}\otimes\ket{\psi_{N-1}}$, then at this time $t$, for the separable qubit, one will again find $\bm{w}\cdot\bm{w}=1$.

Even if $U[t-t_i]\ket{i}$ is completely entangled, there is another special case in which one recovers this equation.  Instead of only looking to the standard forward-evolved state, one must also consider the retro-evolved state $U^{-1}[t_f-t]\ket{f}$.  If this state is partially separable (even if the standard state vector is not) one will again find that $\bm{w}\cdot\bm{w}=1$.  This (hyperbolic) normalization condition sometimes holds in other cases, even when neither state vector is separable.  An explicit example of such a case will be given in Section V.B.

Another special result occurs when two qubits undergo an exchange interaction.  Consider some time $t$ where both qubits are separable in each of the state vectors (both $U[t-t_i]\ket{i}$ and $U^{-1}[t_f-t]\ket{f}$).  In this case, if the weak values $\bm{w_a}$ and $\bm{w_b}$ are interacting in a two-qubit exchange interaction, we find the relationship
\begin{eqnarray}
   Re \frac{d\bm{w_a}}{d\tau}\! &=& \! \frac{\pi}{2} \left[ Re(\!\bm{w_b}\!)
   \!\times \!Re(\!\bm{w_a}\!)\!-\!Im(\!\bm{w_b}\!)\!\times \!Im(\!\bm{w_a}\!) \right]\nonumber \\
   Im \frac{d\bm{w_a}}{d\tau}\! &=& \!\frac{\pi}{2} \left[ Re(\!\bm{w_b}\!)\!\times \!Im(\!\bm{w_a}\!)\!+\!Im(\!\bm{w_b}\!)\!\times \!Im(\!\bm{w_a}\!) \right]. \nonumber
\end{eqnarray}
Furthermore, since $\bm{w}_{avg}$ is constant, the derivatives of $\bm{w}_b$ are just the opposite of $\bm{w}_a$, and can be found by interchanging $\bm{w}_a$ and $\bm{w}_b$.  More compactly, these same equations can be written using a complex cross product,
\begin{eqnarray}
\label{eq:wcross}
    \frac{d\bm{w}_a }{d\tau} &=& \frac{\pi}{2} \bm{w}_b\times\bm{w}_a\label{eq:dwa}\\
    \frac{d\bm{w}_b}{d\tau}  &=& \frac{\pi}{2} \bm{w}_a\times\bm{w}_b\label{eq:dwb}.
\end{eqnarray}
Again, no complex conjugates are used in this calculation.

One could use these equations, for instance, if two qubits were independently prepared, then sent into a SWAP gate, and then independently measured.  At both the beginning and the end of the SWAP gate, the initial time-derivatives of the weak values could be found from the above expressions.  With these in hand, it becomes possible to predict the full evolution through the gate using the second-order equations (\ref{eq:eoms}).  Finding these first derivatives in a more general situation appears to be a much more difficult problem, although a special case where the above equations hold for a non-separable state will be shown in Section V.B.

\subsection{Assessment}

Leaving aside for a moment the question of how the local weak values are \textit{calculated} -- whether using state vectors or some combination of the relationships found in the previous sections -- we have shown that they behave simply and reasonably in any given run of any given quantum circuit.  By analyzing a set of universal quantum gates (the $\sqrt{SWAP}$ and single-qubit rotations), these results immediately generalize to any quantum circuit.

Even more important than reasonable behavior is that the weak values are \textit{localized}, describing events in space and time (rather than in configuration space).  And, unlike the localized particle positions in Bohmian mechanics \cite{norsen}, here there is no evidence of any instantaneous connection between distant events.  Calculating the weak values is found to naturally decouple distant gates from the local dynamics, even if they operate on the same entangled state vector.  Because the precise relative timing is irrelevant, there is reason to hope that a hidden-variable model built on $\bm{w}$ might generally respect Lorentz covariance.

Even for qubits that are not physically localized, as in the case of which-way entanglement (with $\ket{0}$ representing a particle on one path and $\ket{1}$ a particle on another), a localized account is still available.  All that is required, as described in \cite{leifer}, is to assign a localized qubit to \textit{each} path, where $\ket{1}$ is a particle, and $\ket{0}$ is no particle (the ``vacuum state’’ from quantum field theory).  This transforms the original nonlocalized single qubit $a\ket{0}+b\ket{1}$ into \text{two} localized qubits in a larger space, of the form $a\ket{10}+b\ket{01}$.\footnote{This step is likely also necessary in order to make sense of anomalous (negative) weak values; as shown in \cite{wharton2018}, such values can naturally be viewed as indicating that the value has dropped below the typical (non-zero) ``vacuum state’’.}  Such a framework admits two other evident possibilities: $\ket{11}$ (a particle on each path), and $\ket{00}$ (no particle on either path).

Before moving on to discuss eliminating the state vectors from the analysis, we will note what we have already achieved by analyzing the local weak values.  Specifically, a localized physical reality, residing in spacetime rather than configuration space, has been shown to be a possible description for any N-qubit entangled state.  This picture is seen to be consistent with Lorentz covariance despite the spacelike nonlocality apparently suggested by Bell's theorem. At this point, state vectors are still present in the model, but one might argue that they are just useful mathematical tools rather than any essential feature which needs interpretation.

But despite this promising framework, if it proves impossible to eliminate the state vectors from the rules which generate the weak values, then adding weak values as an element of reality would merely be an \textit{extension} of quantum mechanics, not a reformulation. Given the problematic nature of configuration-space wavefunctions, especially as they interface with ordinary space and time, we now turn to an effort to eliminate the state vectors entirely.

\section{Example Model Framework}

The above results motivate a variety of possible models which might locally underpin a generic quantum circuit, as described in the Introduction.  The purpose of this section is to provide a template of what such models might look like, and show some preliminary successes as applied to a partially-entangled state.  Both the models and the suggested extensions are therefore speculative, but well worth considering, if the goal of eliminating state vectors from the analysis is to be taken seriously.

First, consider the possibility that the complex vector $\bm{w}(t)$ for each qubit in a quantum circuit might be determined without using state vectors at all.  The idea is that these vectors could then exactly match the weak values as determined by the standard technique from Eq. (\ref{eq:weak}), and would play the role of the realistic properties $q_{ij}$ on each wire in Figure 1.  This result has already been found to be possible for single qubits in Sections II and III.  If it could also work for multi-qubit circuits, then one could replace all state vectors with the rules that generate $\bm{w}(t)$.  Unfortunately, the fact that $\bm{w}$'s obey second-order (rather than first-order) differential equations in a $\sqrt{SWAP}$ gate makes it difficult to devise such rules.

Another model idea would be to take a different complex 3-vector $\bm{s}(t)$ as the full underlying description for each qubit, but not demand that $\bm{s}=\bm{w}$ in every instance.  Instead, one could consider various solutions $\bm{s}_i(t)$ for any given case, each with an associated probability $P_i$, such that the corresponding weak values $\bm{w}(t)$ were recovered in the statistical limit:
\begin{equation}
\label{eq:probs}
    \sum_i P_i \bm{s}_i(t) = \bm{w}(t).
\end{equation}
This would be consistent with the experimental fact that weak values can only be measured on average. 

In such a statistical model, $\bm{w}$ would be a \textit{guide} to the underlying localized variables, but the actual behavior of $\bm{s}$ could follow different rules in some cases.  For example, even though we found that $\bm{w}\cdot\bm{w}=1$ only in some special circumstances, it would become possible to impose $\bm{s}\cdot\bm{s}=1$ at all times if $\bm{w}$ were a statistical average of the possible values of $\bm{s}$.

For both of these options, we know that $\bm{w}_a(t)$ and $\bm{w}_b(t)$ always obey the equations of motion (\ref{eq:eoms}) whenever they are undergoing an exchange interaction with each other.  Even if one attempts to use the statistical approach, it seems likely that each possible corresponding $\bm{s}_a$ and $\bm{s}_b$ will also have to obey the same equations in every instance:
\begin{eqnarray}
\label{eq:seoms}
    \frac{d^2\bm{s_a}}{d\tau^2} &=&   \frac{\pi^2}{2} (\bm{s_b}-\bm{s_a})\\
    \frac{d^2\bm{s_b}}{d\tau^2} &=&   \frac{\pi^2}{2} (\bm{s_a}-\bm{s_b})\nonumber
\end{eqnarray}

But even with constraints on each preparation and measurement gate, this still leads to a globally underconstrained problem; the first derivatives also need to be determined in some manner.  As we saw in Section IV.C, general first-order equations were difficult to find for $\bm{w}$, but Eq. (\ref{eq:wcross}) did work well in special cases.  For the statistical model, there is no obvious reason why one could not impose these equations on $\bm{s}$.  This strategy will be pursued below. 

A third strategy would be to find some still deeper hidden variable model, such that even $\bm{s}(t)$ would emerge from more fundamental all-at-once rules (ideally a Lagrangian).  This seems like the best long-term approach, but any such analysis would require substantial development in its own right.  That idea will instead be explored in a subsequent publication. 

For now, we will take the simpler and straightforward approach of treating each qubit as a complex 3-vector $\bm{s}$ and guessing some additional constraints which lead to promising preliminary results.  Some problems with this approach will be noted below, but at the very least, this section should be seen as a template for what a future model might eventually look like, with an explicit localized account of every wire in a generic quantum circuit.  

\subsection{Constraints of the Model}

We propose that each localized qubit (each ``wire'' in a quantum circuit) fundamentally consists of a complex 3-vector $\bm{s}$, constrained at all times by the hyperbolic normalization condition $\bm{s}\cdot\bm{s}=1$.  Wherever it is prepared or measured to be aligned with $\hat{\bm{n}}$ (on the Bloch sphere), a constraint imposes $\hat{\bm{n}}\cdot\bm{s}=1$ at that point.  (Notice that this does not set $\bm{s}=\hat{\bm{n}}$, but merely constrains the measured component of $\bm{s}$.)  

When two qubits $\bm{s}_a$ and $\bm{s}_b$ undergo an exchange interaction, other constraints must come into play.  As above, we will normalize the strength of the interaction such that $\tau=1$ corresponds to the SWAP time.  One set of interesting and consistent constraints is as follows:
\begin{eqnarray}
    \bm{s}_a\cdot\bm{s}_a&=&1\label{eq:aa1}\\
    \bm{s}_b\cdot\bm{s}_b&=&1\label{eq:bb1}\\
    \bm{s}_a\cdot\bm{s}_b&=&1\label{eq:ab1}\\
    \frac{d\bm{s}_a }{d\tau} &=& \frac{\pi}{2} \bm{s}_b\times\bm{s}_a\label{eq:dsa}\\
    \frac{d\bm{s}_b}{d\tau}  &=& \frac{\pi}{2} \bm{s}_a\times\bm{s}_b\label{eq:dsb}
\end{eqnarray}

The first two equations follow from the $\bm{s}\cdot\bm{s}=1$ constraint, and the last two equations are motivated by certain results in the previous section.  These first-order differential equations imply the stability of the first three constraint equations (\ref{eq:aa1}-\ref{eq:ab1}).  If they are true at one time, they remain true at all times.  

The new Eq. (\ref{eq:ab1}) is motivated by taking another $\tau$-derivative of (\ref{eq:dsa}) and (\ref{eq:dsb}), which can be combined to yield
\begin{eqnarray}
    \frac{d^2\bm{s_a}}{d\tau^2} &=& \frac{\pi^2}{4} [\bm{s}_a\cdot\bm{s}_a+\bm{s}_a\cdot\bm{s}_b](\bm{s}_b -\bm{s}_a) \label{eq:ddsa}\\
    \frac{d^2\bm{s_b}}{d\tau^2} &=& \frac{\pi^2}{4} [\bm{s}_b\cdot\bm{s}_b+\bm{s}_a\cdot\bm{s}_b] (\bm{s}_a - \bm{s}_b)\label{eq:ddsb}.
\end{eqnarray}
If Eqs. (\ref{eq:aa1}-\ref{eq:ab1}) are enforced, the nonlinear aspect of these equations disappears; the bracketed terms on the right-hand side become $2$, and one recovers (\ref{eq:seoms}) exactly.  Therefore, this model does not need to impose those second-order equations as additional constraints; the required dynamics follows automatically from the above list.  

Of all these constraints, perhaps the most surprising is Eq. (\ref{eq:ab1}), which is mathematically well defined and consistent, but still counterintuitive.  It says that as one brings two qubits together into an exchange interaction, their (hidden) complex vectors are already correlated with each other even before the interaction begins.  Evidently, there is no forward-causal account of how this might be arranged.  But recall that this entire research program is predicated on solving these circuits ``all at once'', rather than sequentially.  In such a framework, this ``pre-arrangement'' of the hidden qubit states is both expected and required.  And it is not the only constraint with this feature; imposing $\hat{\bm{n}}\cdot\bm{s}=1$ at future measurements will also have consequences at earlier times.  Regardless of the interpretation, we can proceed to impose these constraints mathematically and then see what happens inside a particular circuit.  

\subsection{Weak Values of a $\sqrt{SWAP}$ Interaction}

Consider the simple quantum circuit in Figure 4, a single $\sqrt{SWAP}$ gate.  The input states have been prepared to point in the $+\hat{\bm{x}}$ and $+\hat{\bm{y}}$ directions.  A conventional QM calculation would indicate that the initial state was therefore
\begin{eqnarray}
    \ket{i}&=&\frac{1}{\sqrt{2}}(\ket{0}+\ket{1})\otimes \frac{1}{\sqrt{2}}(\ket{0}+i\ket{1}) \nonumber \\
    &=& \frac{1}{2} (\ket{00}+i\ket{01}+\ket{10}+i\ket{11}).
\end{eqnarray}

The effect of the $\sqrt{SWAP}$ gate is then to create the partially entangled state $\ket{\psi_f}=U_{ex}[0.5]\ket{i}$, which can be computed from Eq. (\ref{eq:Uex}) as
\begin{equation}
    \ket{\psi_f}= \frac{1}{2}\ket{00}+\frac{1+i}{2}\ket{10}+\frac{i}{2}\ket{11}.
\end{equation}
An application of the Born rule then indicates that the probability of the $\ket{00}$ and $\ket{11}$ outcomes are 25\% each, and the probability of the $\ket{10}$ outcome is 50\%.  The fourth outcome has zero probability; it is impossible to measure the first qubit in the $+\hat{\bm{z}}$ direction (corresponding to $\ket{0}$) while also measuring the second qubit in the $-\hat{\bm{z}}$ direction (corresponding to $\ket{1}$). 

\begin{figure}[htbp]
\begin{center}
\includegraphics[width=6cm]{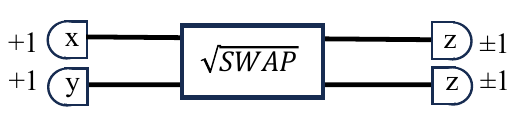}
\caption{The central example considered in this section is a $\sqrt{SWAP}$ gate operating on a separable state $\ket{i}=\hat{\bm{x}}\otimes\hat{\bm{y}}$, with both outcomes measured in the $(\ket{0},\ket{1})$ basis, corresponding to Bloch vectors $(\hat{\bm{z}},-\hat{\bm{z}})$, and outcomes $(+1,-1)$.  A final measurement of $\ket{00}$ would correspond to both classical outcomes being equal to $+1$, but other outcomes are also possible.}
\label{default}
\end{center}
\end{figure}

Although this is a simple calculation, it provides no physical insight as to what might be happening inside the $\sqrt{SWAP}$ gate. As far as spacetime-based variables are concerned, it is a literal black box.  The computed output state $\ket{\psi_f}$ is not separable, so it is impossible to say what is happening on each individual output wire.  There is also no physical or intuitive account of why one outcome is impossible, and why the other three outcomes have different probabilities.  

But, as we saw in the above sections, additional details can be found by looking at $\bm{w}$.  For any possible measured eigenstate $\ket{f}$, the weak values can be computed exactly, for each qubit at every moment, with full analytical results $\bm{w_a}(\tau)$ and $\bm{w_b}(\tau)$ given in Appendix B.  To get a general sense of these results, the initial and final values of $\bm{w}_a$ and $\bm{w}_b$ are shown in Figure 5 for each possibility.

\begin{figure}[htbp]
\begin{center}
\includegraphics[width=8cm]{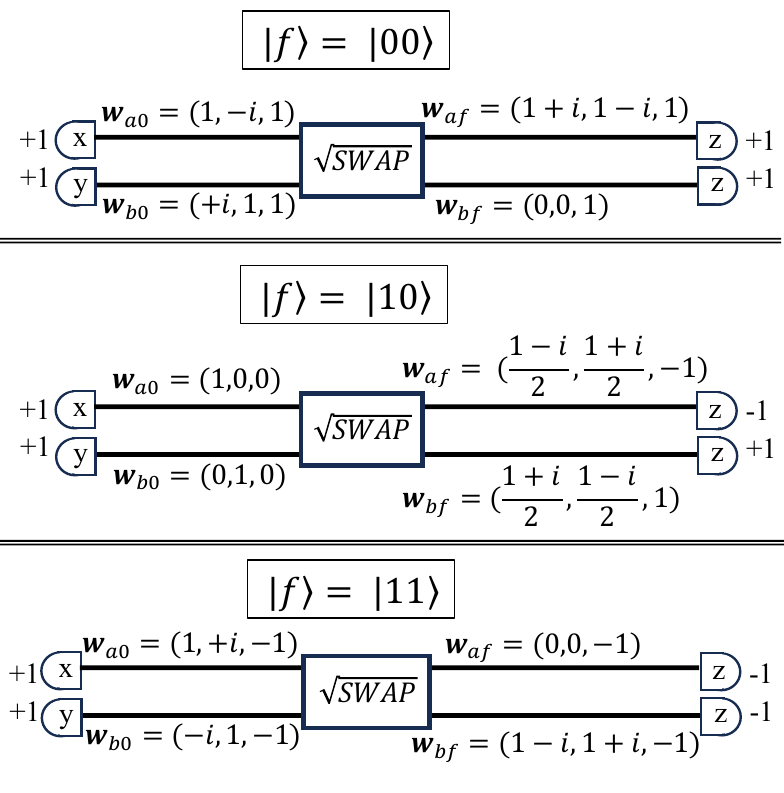}
\caption{The local weak values $\bm{w}$ are shown for the $\sqrt{SWAP}$ circuit from in this section, with both initial and final weak values labeled on the circuit itself. (Full results are in Appendix B.)  Each of the three outcomes with non-zero probability are shown; the weak values do not exist for the fourth.}
\label{default}
\end{center}
\end{figure}

Looking at the full history $\bm{w}(\tau)$, two of the three cases shown in Figure 5 reveal an intriguing pattern.  For both the $\ket{f}=\ket{00}$ and $\ket{f}=\ket{11}$ outcome (25\% probabilities), the single-qubit conditions $\bm{w}_a\cdot\bm{w}_a=1$ and $\bm{w}_b\cdot\bm{w}_b=1$ hold true at every moment.  Furthermore, one also finds $\bm{w}_a\cdot\bm{w}_b=1$ at every point.  Even more surprisingly, Eqs. (\ref{eq:dwa}) and (\ref{eq:dwb}) are correct at every intermediate point in these particular solutions.  In other words, the solutions for these two cases already obey the model constraints Eqs. (\ref{eq:aa1}-\ref{eq:dsb}), with $\bm{s}\to\bm{w}$.

But all of these patterns fail when analyzing the solution with the $\ket{10}$ outcome (50\% probability).  At the very beginning and the very end of that circuit, one sees from the middle of Figure 5 that $\bm{w}_a\cdot\bm{w}_a=1$ and $\bm{w}_b\cdot\bm{w}_b=1$, but this is only because the one of the two state vectors is separable at that point.  At any other time, these equations do not hold true for this particular solution.  Also, $\bm{w}_a\cdot\bm{w}_b\ne 1$, and the first derivatives $\dot{\bm{w}}$ are never predicted by Eqs. (\ref{eq:wcross}).  

It is a serious problem to have a set of constraint equations that work in one instance but not in another.  One solution is to ignore the precise behavior of $\bm{w}$ and focus on lower-level rules that constrain $\bm{s}$.  We will now show that maintaining consistent constraints on $\bm{s}$ can recover the average behavior of $\bm{w}_a$ and $\bm{w}_b$, even for the problematic $\ket{f}=\ket{10}$ case.

\subsection{An Improved Model of the $\sqrt{SWAP}$ Interaction}

Although it was difficult to make sense of $\bm{w}$ in the $\sqrt{SWAP}$ gate (at least for the $\ket{f}=\ket{10}$ outcome), this circuit works much better when analyzed in terms of $\bm{s}$.  This requires not only imposing the constraints from Eqs. (\ref{eq:aa1}-\ref{eq:dsb}), but also four additional constraints from the measurement and the preparation.  In terms of the initial vectors $\bm{s}_{a0}$, $\bm{s}_{b0}$ and the final vectors $\bm{s}_{af}$, $\bm{s}_{bf}$, these constraints are
\begin{eqnarray}
    \hat{\bm{x}}\cdot \bm{s}_{a0}&=&1\\
    \hat{\bm{y}}\cdot \bm{s}_{b0}&=&1\\
    \pm_a\hat{\bm{z}}\cdot \bm{s}_{af}&=&1\\
    \pm_b\hat{\bm{z}}\cdot \bm{s}_{bf}&=&1.
\end{eqnarray}
The signs in the last two equations correspond to the measured direction, with the positive sign corresponding to $\ket{0}$, and the negative sign corresponding to $\ket{1}$.  

Choosing a pair of signs in these last equations, we started with randomly generated seeds and used a gradient-descent technique (BRST algorithm) to numerically search for solutions for these 4 constraint equations, along with the further constraints in Eqns. (\ref{eq:aa1}-\ref{eq:dsb}).  All generated solutions were simple enough to check analytically, so we can be confident they are exact.  These results are shown in Figure 6.   

\begin{figure*}[htbp]
\begin{center}
\includegraphics[width=16cm]{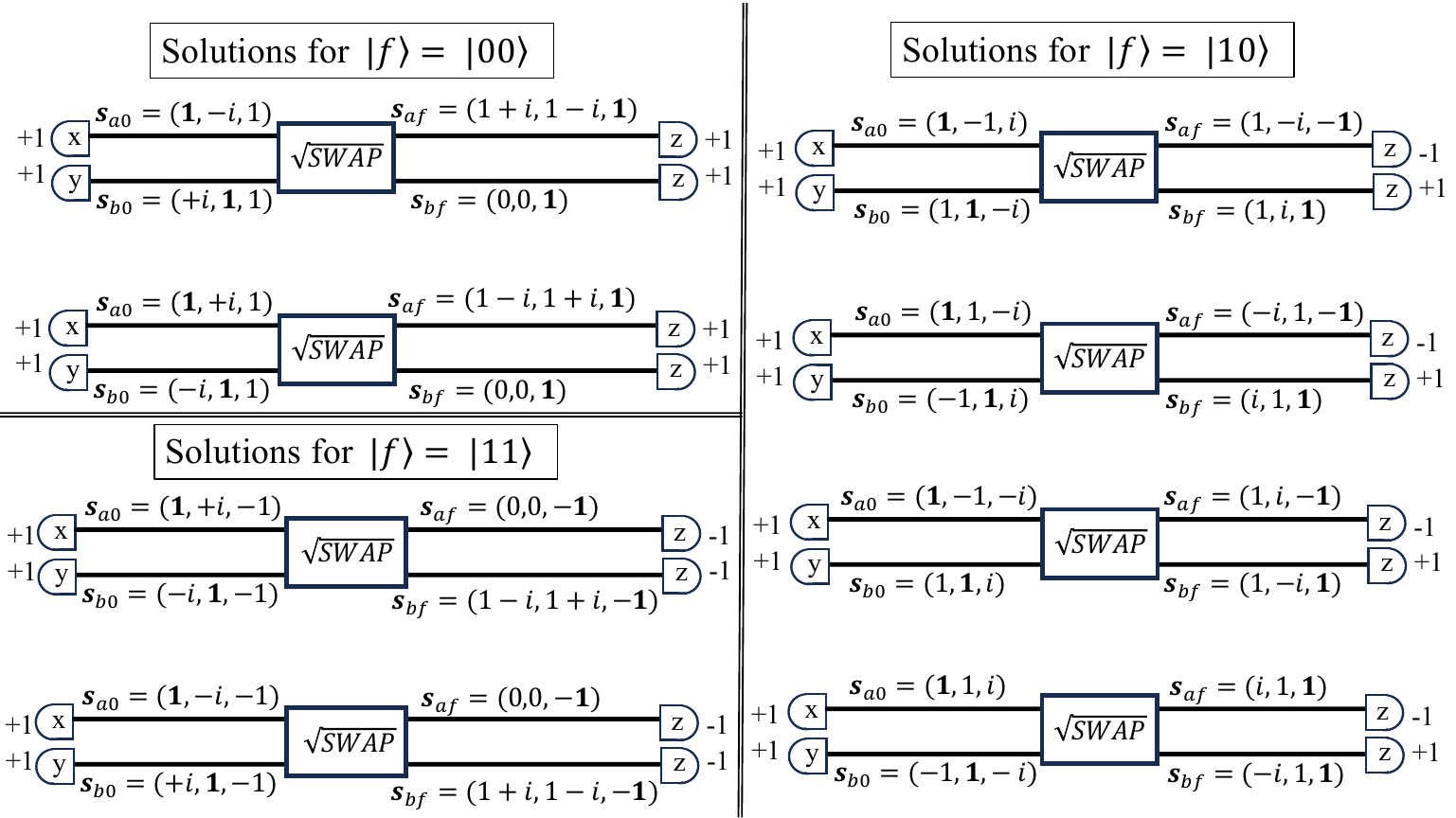}
\caption{The comprehensive set of solutions for the $\sqrt{SWAP}$ circuit from in this section, with both initial and final 3-vectors labeled on the circuit itself.  The values matching the preparation and measurement constraints are shown in bold.  Two distinct solutions exist for each of the $\ket{00}$ and $\ket{11}$ outcomes; four solutions exist for $\ket{f}=\ket{10}$.  No solutions exist for $\ket{f}=\ket{01}$.}
\label{default}
\end{center}
\end{figure*}

It should be evident from Figure 6 that the number of solutions exactly match the probabilities of the outcomes.  Each of the 25\% probability outcomes ($\ket{00}$ and $\ket{11}$) has two solutions, the 50\% probability outcome ($\ket{10}$) has four, and the impossible outcome has no solutions at all.  This is a promising result, to be discussed below. 

But this is not a mere black-box calculation; it is also possible to look inside the interaction to see why these solutions might make sense.  For any given solution, at every point in time inside the $\sqrt{SWAP}$ gate, the qubits $\bm{s}_a$ and $\bm{s}_b$ are evolving according to (\ref{eq:dsa}) and (\ref{eq:dsb}).  In every case, the preparation sets the $x$-component of $\bm{s}_a$ equal to 1, along with the $y$-component of $\bm{s}_b$. From Eq. (\ref{eq:dsa}), the $z$-component of the $\tau$-derivative of $\bm{s}_a$ is proportional to $(\bm{s}_b\times\bm{s}_a)\cdot\hat{\bm{z}}$.  (If one considers $\bm{s}$ to be a spin vector, then this derivative is a ``torque''.)  There is not enough information from only the preparation to solve for the torque exactly, but from what we know about the $x$ and $y$ components, the initial $z$-directed torque of $\bm{s}_a$ will tend to be negative.  Similarly, the $z$-directed torque of $\bm{s}_b$ will tend to be positive.  Therefore the torques tend to push $\bm{s}_a$ in the negative-$z$ direction, and we find more solutions when this is the case.  The opposite is true for $\bm{s}_b$.  And if one attempts to run them both in the ``wrong'' direction, imposing a final boundary constraint corresponding to a positive $\bm{s}_a\cdot\hat{\bm{z}}$ and a negative $\bm{s}_b\cdot\hat{\bm{z}}$, it is not surprising that there are no solutions at all.

Another intriguing feature of these results is that the original $Re(\bm{w})$ can be generated by averaging $Re(\bm{s})$ for each particular case, including the special $\ket{10}$ outcome.  This averaging procedure works not only at the endpoints, but also at every intermediate value of $\tau$.  It is a point in favor of this example model that the weak values re-emerge in this manner.  After all, to the extent that weak values can be measured in the lab, they can only be measured on average, so there is no reason an underlying model should not have precisely this same feature.

The imaginary part of the weak value is not recovered on average, unless half of the solutions are eliminated.\footnote{If the last half of each set of solutions in Figure 6 are discarded, then the average $<Im(\bm{s})>$ does indeed equal $Im(\bm{w})$.}  As the complex conjugate of any solution is also a solution, this symmetry would have to be broken for any non-zero average.  Another option would be to interpret $Im(\bm{w})$ as telling us something about the variance of $Im(\bm{s})$, rather than $Im(\bm{s})$ itself, but it would be premature to address this detail before turning to more serious problems.

\subsection{Model Discussion}

There are several problems with this example model, of different levels of severity. The probability rule of counting the number of solutions can clearly not be generalizable to other cases, at least not with a small finite solution number.  But this was already evident from the single-qubit sector, where it was found that the probability of single-qubit outcomes scaled like $1/Re[\bm{w}]^2$, without counting solutions at all.  It would not be difficult to combine two rules along these lines; weighting each solution by a relative probability $P_i$ based on some generalized expression of this sort.  

For this example, it is worth noting that having twice as many solutions did not lead to four times the probability.  This is not a path-integral calculation, where one adds amplitudes and squares the result.  Instead, it is more like a calculation in classical statistical mechanics, where one literally counts the number of possibilities to find a relative likelihood.  

Still, it is not clear what probability rule would work for both this example and for the single-qubit case.  Looking at the solutions $\bm{s}(\tau)$, it is clear that they all begin with the same magnitude $Re[\bm{s}]$, but that similarity starts to disappear in the middle of the exchange interaction.  By the time the qubits stop interacting, the resulting values $\bm{s}_{af}$ and $\bm{s}_{bf}$ look somewhat different for the $\ket{00}$ and $\ket{11}$ outcomes than they do for the $\ket{10}$ case.  Even given this difference, the average of $Re[\bm{s}_{af}]^2$ and $Re[\bm{s}_{bf}]^2$ is identical for these three cases, making it plausible that the total probability might simply scale with the number of solutions for this one example.

A better probability rule could not be proposed without first tackling the biggest problem: over-constrained solutions.  Counting real numbers, the initial state of a 2-qubit system has 12 real variables, 6 for each complex 3-vector.  Since they solve first-order differential equations (\ref{eq:dsa}) and (\ref{eq:dsb}), these initial 12 values determine the full history.  Constraining these 12 values, the model has three complex equations, (\ref{eq:aa1}), (\ref{eq:bb1}), and (\ref{eq:ab1}).  Since the the left side of these equations is generally complex, the right side should be read as $1+0i$, meaning that each of these equations provides two real constraints, for 6 total.  Similarly, each preparation and measurement constraint $\hat{\bm{n}}\cdot \bm{s}=1+0i$ provides two more real constraints, for 8 more total.  This is 14 constraints on 12 variables, so the fact that solutions were available for this example turns out to be a special case.

The simplest resolution to this problem is to relax the preparation/measurement constraint in the imaginary sector, changing it to $\hat{\bm{n}}\cdot Re[\bm{s}]=1$.  This would be 10 constraints on 12 variables.  Preliminary work along these lines indicates that the most promising approach would fold the previous constraint $\hat{\bm{n}}\cdot Im[\bm{s}]=0$ into a probability rule.  In other words, one would enforce all 14 constraints whenever possible, but still allow four of these constraints to be relaxed in at least two places if needed.

An even more promising resolution would be to expand the hidden variable space for each qubit. Evidently, 14 constraints would still be acceptable if the model required 7 or 8 real variables per qubit, instead of 6.  A development of this idea will appear in a future publication, based on a Clifford-algebra version of the ``second-order qubits'' described in \cite{wharton2015}.

\section{Summary and Conclusions}

\subsection{Local Weak Values as Elements of Reality}

Following Einstein \cite{einstein1935}, John Bell began his analysis of entanglement experiments by trying to find an account in terms of ``local beables'' (be-ables, as distinct from observe-ables), objectively existing even in the absence of measurement.  This requirement that a deeper description of entangled states should concern real events localized in space and time seemed obvious to both him and Einstein.  His subsequent no-go theorem was sometimes interpreted as ruling out the existence of local beables, but that was never how Bell saw the situation.  He clearly saw that any precise reformulation of QM would likely concern \textit{localized} beables, still associated with locations in space and time, but with properties that violated one of the assumptions of his theorem.  \cite{bell2004}

Section II began by looking to the local weak values in any complete quantum circuit as a possible candidate for Bell's localized beables.  They are seen to be promising in this regard, not only because they can be associated with any particular wire in any quantum circuit, but also because they do not grow exponentially with the number of qubits elsewhere in the circuit.

Pursuing this analysis, the results in Section III then demonstrated the surprising fact that these weak values exhibit dynamic locality, remaining constant on each circuit wire regardless of what happens elsewhere.  The corresponding weak value vectors also rotate as expected through single-qubit gates, even if they are generated from a massively-entangled state. 

Section IV demonstrated that in the middle of a universal two-qubit gate, the weak values of the qubits evolve according to a local oscillator interaction, regardless of what is currently happening elsewhere.  And since generic quantum circuits can be built from this basic gate set, the above results can be read as applying to any quantum circuit whatsoever.

The reason these results are surprising is the widespread view that violations of Bell inequalities in quantum circuits are due to some sort of direct influence or connection between spatially-separated parts of the circuit.  Certainly, if this were happening to the weak values, it would explain how the Bell inequality violations would be achieved.  But evidently that is not what is happening; the weak values reveal no appearance of any simultaneous link between distant events.  Although the Bell inequalities are indeed broken, the local weak values must be violating Bell's assumptions in some other way.

And indeed, weak values can be seen to instead violate the ``statistical independence'' assumption which goes into the proofs of Bell/CHSH inequalities.  Even in the single-qubit examples from Section II, the allowable weak values depend upon the future basis in which the qubit is measured.  Such ``future-input-dependent'' models have been highlighted as a promising way to explain Bell inequality violations while retaining the central elements needed for relativistic invariance.\cite{wharton2020}  The fact that analysis of the local weak values leads to precisely this sort of model could be seen as further evidence that this is a plausible way to better describe what might be really happening between measurements.

The results through Section IV therefore comprise several strands of evidence for a realistic and localized underpinning of any circuit, as described in the Introduction.  Already, this motivates a natural extension of quantum theory to include these local weak values as an additional element of reality.  But for many researchers in quantum foundations, the goal is not merely to extend QM, but to reformulate the theory in a way that provides new explanatory power.  With this point in mind, we turn to the successes and failures of the model presented in Section V.

\subsection{Towards a Localized Reformulation of Quantum Circuits}

The model described in Section V addresses the possibility that an eventual underpinning of a generic quantum circuit might have to drop to a level more fundamental than the weak values.  In the $\sqrt{SWAP}$ circuit, imposing the $\ket{f}=\ket{10}$ outcome led to two weak value vectors ($\bm{w}_a$ and $\bm{w}_b$) which did not obey the same rules as the vectors for the other allowable outcomes.  It is evidently difficult to build a model with one set of rules for one possible solution, but other rules for a different possible solution.

The results in Section V.C show that this problem can be overcome by setting aside the actual weak value vector $\bm{w}$ and postulating a different complex vector $\bm{s}$ to represent the hidden structure of each qubit.  We then imposed a consistent set of constraints on $\bm{s}$.  For this purpose, we chose constraints that were sometimes, but not always, successful for $\bm{w}$.  We then expect the solutions for $\bm{s}$ to generally deviate from the weak values $\bm{w}$.

The surprising result of this analysis was that $Re[\bm{w}]$ was still recovered on \textit{average}, from all the allowable solutions of $Re[\bm{s}]$.  If a symmetry between conjugate solutions is broken in a certain way, the imaginary parts of these vectors also matched each other.  This indicates that the occasionally-unpredictable behavior of $\bm{w}$ may not be a critical problem, but rather an indication that the weak values are inherently averaging over lower-level behavior. 

Even more promising, the probabilities for the possible outcomes matched the size of the $\bm{s}$ solution space, providing an intuitive reason for their relative likelihoods.  True, this assumed that every solution was equally probable, which seems unlikely to hold when a wider variety of circuits is considered.  But the correct (unequal) probabilities were already found to be generated by Eq. (\ref{eq:prob1}) for the case of a single qubit, and there are several options for generalizing this expression.

The plausibility of eventually finding a successful probability rule can also be seen from the original definition of the weak value.  Looking back at Eq. (\ref{eq:weak}), one sees that the denominator of the weak value, if squared, becomes the probability of that particular outcome.  So, in at least some sense, the probabilities are already encoded in $\bm{w}$; any procedure that can generate $\bm{w}$ -- even on average -- would likely be able to supply an additional rule to generate the outcome probabilities.  Again, the state vectors are seen to be potentially discardable, given a replacement model that can generate the weak values on average.

Despite all these successes, the model from Section V should merely be read as a template for localized models of quantum entanglement, as it is not yet able to replace state vectors entirely.  The biggest problem was that it utilized too many constraints, given the number of variables.  Despite this drawback, it was still able to use spacetime-localized variables to successfully model a situation which normally requires a partially-entangled state vector.  To our knowledge this has never been done before, without also using state vectors themselves at some point in the analysis.  Previous future-input-dependent toy models (as surveyed in \cite{wharton2020}) have only been able to model maximally-entangled states (with a recent extension to GHZ states \cite{neder2024}).  Localized models of partially-entangled states have previously only been able to explain the zero-probability cases \cite{whartonadlam}, a benchmark that the Section V model easily surpassed.

There is reason to hope that this style of incremental progress, modeling one type of entangled state at a time, is nearing its end.  The promise of framing these models in terms of quantum circuits is that they would then become as modular as the circuits themselves.  Only a few basic rules for a few basic gates would suffice to cover all possible situations.  If someone is able to develop a successful general model for a $\sqrt{SWAP}$ gate, in a manner compatible with the single-qubit results of Section III, then an infinite number of entanglement geometries could be analyzed using this modular strategy. 

Given a large number of possible ways to modify the Section V model, it seems plausible that a fully successful model can be found, using the weak values as a guide.  The fact that the weak values are seen to evolve locally, even in massively entangled states, is evidence that this is a reasonable approach.  Dressel has noted that if a model ``could really mimic the detailed functional structure of the weak value, then it would also be able to simulate other features that are normally considered to be quantum mechanical’’ \cite{dressel2015}.  If this could indeed be achieved as outlined above, at the mere expense of using measurement settings as an explanatory resource, then a concerted push to find such a reformulation should arguably be a central priority for modern research in quantum foundations.

\begin{acknowledgments}
The authors gratefully thank Nathan Argaman for helpful feedback.
\end{acknowledgments}

\bibliography{References}

\begin{thebibliography}{28}%
\makeatletter
\providecommand \@ifxundefined [1]{%
 \@ifx{#1\undefined}
}%
\providecommand \@ifnum [1]{%
 \ifnum #1\expandafter \@firstoftwo
 \else \expandafter \@secondoftwo
 \fi
}%
\providecommand \@ifx [1]{%
 \ifx #1\expandafter \@firstoftwo
 \else \expandafter \@secondoftwo
 \fi
}%
\providecommand \natexlab [1]{#1}%
\providecommand \enquote  [1]{``#1''}%
\providecommand \bibnamefont  [1]{#1}%
\providecommand \bibfnamefont [1]{#1}%
\providecommand \citenamefont [1]{#1}%
\providecommand \href@noop [0]{\@secondoftwo}%
\providecommand \href [0]{\begingroup \@sanitize@url \@href}%
\providecommand \@href[1]{\@@startlink{#1}\@@href}%
\providecommand \@@href[1]{\endgroup#1\@@endlink}%
\providecommand \@sanitize@url [0]{\catcode `\\12\catcode `\$12\catcode `\&12\catcode `\#12\catcode `\^12\catcode `\_12\catcode `\%12\relax}%
\providecommand \@@startlink[1]{}%
\providecommand \@@endlink[0]{}%
\providecommand \url  [0]{\begingroup\@sanitize@url \@url }%
\providecommand \@url [1]{\endgroup\@href {#1}{\urlprefix }}%
\providecommand \urlprefix  [0]{URL }%
\providecommand \Eprint [0]{\href }%
\providecommand \doibase [0]{https://doi.org/}%
\providecommand \selectlanguage [0]{\@gobble}%
\providecommand \bibinfo  [0]{\@secondoftwo}%
\providecommand \bibfield  [0]{\@secondoftwo}%
\providecommand \translation [1]{[#1]}%
\providecommand \BibitemOpen [0]{}%
\providecommand \bibitemStop [0]{}%
\providecommand \bibitemNoStop [0]{.\EOS\space}%
\providecommand \EOS [0]{\spacefactor3000\relax}%
\providecommand \BibitemShut  [1]{\csname bibitem#1\endcsname}%
\let\auto@bib@innerbib\@empty
\bibitem [{\citenamefont {Wharton}\ and\ \citenamefont {Argaman}(2020)}]{wharton2020}%
  \BibitemOpen
  \bibfield  {author} {\bibinfo {author} {\bibfnamefont {K.}~\bibnamefont {Wharton}}\ and\ \bibinfo {author} {\bibfnamefont {N.}~\bibnamefont {Argaman}},\ }\bibfield  {title} {\bibinfo {title} {Colloquium: Bell’s theorem and locally mediated reformulations of quantum mechanics},\ }\href@noop {} {\bibfield  {journal} {\bibinfo  {journal} {Reviews of Modern Physics}\ }\textbf {\bibinfo {volume} {92}},\ \bibinfo {pages} {021002} (\bibinfo {year} {2020})}\BibitemShut {NoStop}%
\bibitem [{\citenamefont {Roberts}(1978)}]{roberts1978}%
  \BibitemOpen
  \bibfield  {author} {\bibinfo {author} {\bibfnamefont {K.}~\bibnamefont {Roberts}},\ }\bibfield  {title} {\bibinfo {title} {An objective interpretation of lagrangian quantum mechanics},\ }\href@noop {} {\bibfield  {journal} {\bibinfo  {journal} {Proceedings of the Royal Society of London. A. Mathematical and Physical Sciences}\ }\textbf {\bibinfo {volume} {360}},\ \bibinfo {pages} {135} (\bibinfo {year} {1978})}\BibitemShut {NoStop}%
\bibitem [{\citenamefont {Aharonov}\ \emph {et~al.}(1988)\citenamefont {Aharonov}, \citenamefont {Albert},\ and\ \citenamefont {Vaidman}}]{AAV}%
  \BibitemOpen
  \bibfield  {author} {\bibinfo {author} {\bibfnamefont {Y.}~\bibnamefont {Aharonov}}, \bibinfo {author} {\bibfnamefont {D.~Z.}\ \bibnamefont {Albert}},\ and\ \bibinfo {author} {\bibfnamefont {L.}~\bibnamefont {Vaidman}},\ }\bibfield  {title} {\bibinfo {title} {How the result of a measurement of a component of the spin of a spin-1/2 particle can turn out to be 100},\ }\href@noop {} {\bibfield  {journal} {\bibinfo  {journal} {Physical review letters}\ }\textbf {\bibinfo {volume} {60}},\ \bibinfo {pages} {1351} (\bibinfo {year} {1988})}\BibitemShut {NoStop}%
\bibitem [{\citenamefont {Sutherland}(1998)}]{sutherland1998}%
  \BibitemOpen
  \bibfield  {author} {\bibinfo {author} {\bibfnamefont {R.~I.}\ \bibnamefont {Sutherland}},\ }\bibfield  {title} {\bibinfo {title} {Density formalism for quantum theory},\ }\href@noop {} {\bibfield  {journal} {\bibinfo  {journal} {Foundations of Physics}\ }\textbf {\bibinfo {volume} {28}},\ \bibinfo {pages} {1157} (\bibinfo {year} {1998})}\BibitemShut {NoStop}%
\bibitem [{\citenamefont {Campagne-Ibarcq}\ \emph {et~al.}(2014)\citenamefont {Campagne-Ibarcq}, \citenamefont {Bretheau}, \citenamefont {Flurin}, \citenamefont {Auff{\`e}ves}, \citenamefont {Mallet},\ and\ \citenamefont {Huard}}]{exp1}%
  \BibitemOpen
  \bibfield  {author} {\bibinfo {author} {\bibfnamefont {P.}~\bibnamefont {Campagne-Ibarcq}}, \bibinfo {author} {\bibfnamefont {L.}~\bibnamefont {Bretheau}}, \bibinfo {author} {\bibfnamefont {E.}~\bibnamefont {Flurin}}, \bibinfo {author} {\bibfnamefont {A.}~\bibnamefont {Auff{\`e}ves}}, \bibinfo {author} {\bibfnamefont {F.}~\bibnamefont {Mallet}},\ and\ \bibinfo {author} {\bibfnamefont {B.}~\bibnamefont {Huard}},\ }\bibfield  {title} {\bibinfo {title} {Observing interferences between past and future quantum states in resonance fluorescence},\ }\href@noop {} {\bibfield  {journal} {\bibinfo  {journal} {Physical review letters}\ }\textbf {\bibinfo {volume} {112}},\ \bibinfo {pages} {180402} (\bibinfo {year} {2014})}\BibitemShut {NoStop}%
\bibitem [{\citenamefont {Weber}\ \emph {et~al.}(2014)\citenamefont {Weber}, \citenamefont {Chantasri}, \citenamefont {Dressel}, \citenamefont {Jordan}, \citenamefont {Murch},\ and\ \citenamefont {Siddiqi}}]{exp2}%
  \BibitemOpen
  \bibfield  {author} {\bibinfo {author} {\bibfnamefont {S.}~\bibnamefont {Weber}}, \bibinfo {author} {\bibfnamefont {A.}~\bibnamefont {Chantasri}}, \bibinfo {author} {\bibfnamefont {J.}~\bibnamefont {Dressel}}, \bibinfo {author} {\bibfnamefont {A.~N.}\ \bibnamefont {Jordan}}, \bibinfo {author} {\bibfnamefont {K.~W.}\ \bibnamefont {Murch}},\ and\ \bibinfo {author} {\bibfnamefont {I.}~\bibnamefont {Siddiqi}},\ }\bibfield  {title} {\bibinfo {title} {Mapping the optimal route between two quantum states},\ }\href@noop {} {\bibfield  {journal} {\bibinfo  {journal} {Nature}\ }\textbf {\bibinfo {volume} {511}},\ \bibinfo {pages} {570} (\bibinfo {year} {2014})}\BibitemShut {NoStop}%
\bibitem [{\citenamefont {Tan}\ \emph {et~al.}(2015)\citenamefont {Tan}, \citenamefont {Weber}, \citenamefont {Siddiqi}, \citenamefont {M{\o}lmer},\ and\ \citenamefont {Murch}}]{exp3}%
  \BibitemOpen
  \bibfield  {author} {\bibinfo {author} {\bibfnamefont {D.}~\bibnamefont {Tan}}, \bibinfo {author} {\bibfnamefont {S.}~\bibnamefont {Weber}}, \bibinfo {author} {\bibfnamefont {I.}~\bibnamefont {Siddiqi}}, \bibinfo {author} {\bibfnamefont {K.}~\bibnamefont {M{\o}lmer}},\ and\ \bibinfo {author} {\bibfnamefont {K.}~\bibnamefont {Murch}},\ }\bibfield  {title} {\bibinfo {title} {Prediction and retrodiction for a continuously monitored superconducting qubit},\ }\href@noop {} {\bibfield  {journal} {\bibinfo  {journal} {Physical review letters}\ }\textbf {\bibinfo {volume} {114}},\ \bibinfo {pages} {090403} (\bibinfo {year} {2015})}\BibitemShut {NoStop}%
\bibitem [{\citenamefont {Rybarczyk}\ \emph {et~al.}(2015)\citenamefont {Rybarczyk}, \citenamefont {Peaudecerf}, \citenamefont {Penasa}, \citenamefont {Gerlich}, \citenamefont {Julsgaard}, \citenamefont {M{\o}lmer}, \citenamefont {Gleyzes}, \citenamefont {Brune}, \citenamefont {Raimond}, \citenamefont {Haroche} \emph {et~al.}}]{exp4}%
  \BibitemOpen
  \bibfield  {author} {\bibinfo {author} {\bibfnamefont {T.}~\bibnamefont {Rybarczyk}}, \bibinfo {author} {\bibfnamefont {B.}~\bibnamefont {Peaudecerf}}, \bibinfo {author} {\bibfnamefont {M.}~\bibnamefont {Penasa}}, \bibinfo {author} {\bibfnamefont {S.}~\bibnamefont {Gerlich}}, \bibinfo {author} {\bibfnamefont {B.}~\bibnamefont {Julsgaard}}, \bibinfo {author} {\bibfnamefont {K.}~\bibnamefont {M{\o}lmer}}, \bibinfo {author} {\bibfnamefont {S.}~\bibnamefont {Gleyzes}}, \bibinfo {author} {\bibfnamefont {M.}~\bibnamefont {Brune}}, \bibinfo {author} {\bibfnamefont {J.}~\bibnamefont {Raimond}}, \bibinfo {author} {\bibfnamefont {S.}~\bibnamefont {Haroche}}, \emph {et~al.},\ }\bibfield  {title} {\bibinfo {title} {Forward-backward analysis of the photon-number evolution in a cavity},\ }\href@noop {} {\bibfield  {journal} {\bibinfo  {journal} {Physical Review A}\ }\textbf {\bibinfo {volume} {91}},\ \bibinfo {pages} {062116} (\bibinfo {year} {2015})}\BibitemShut {NoStop}%
\bibitem [{\citenamefont {Sutherland}(2022)}]{sutherland2022}%
  \BibitemOpen
  \bibfield  {author} {\bibinfo {author} {\bibfnamefont {R.~I.}\ \bibnamefont {Sutherland}},\ }\bibfield  {title} {\bibinfo {title} {Probabilities and certainties within a causally symmetric model},\ }\href@noop {} {\bibfield  {journal} {\bibinfo  {journal} {Foundations of Physics}\ }\textbf {\bibinfo {volume} {52}},\ \bibinfo {pages} {75} (\bibinfo {year} {2022})}\BibitemShut {NoStop}%
\bibitem [{\citenamefont {Dressel}(2015)}]{dressel2015}%
  \BibitemOpen
  \bibfield  {author} {\bibinfo {author} {\bibfnamefont {J.}~\bibnamefont {Dressel}},\ }\bibfield  {title} {\bibinfo {title} {Weak values as interference phenomena},\ }\href@noop {} {\bibfield  {journal} {\bibinfo  {journal} {Physical Review A}\ }\textbf {\bibinfo {volume} {91}},\ \bibinfo {pages} {032116} (\bibinfo {year} {2015})}\BibitemShut {NoStop}%
\bibitem [{\citenamefont {Sutherland}(2008)}]{sutherland2008}%
  \BibitemOpen
  \bibfield  {author} {\bibinfo {author} {\bibfnamefont {R.~I.}\ \bibnamefont {Sutherland}},\ }\bibfield  {title} {\bibinfo {title} {Causally symmetric bohm model},\ }\href@noop {} {\bibfield  {journal} {\bibinfo  {journal} {Studies in History and Philosophy of Science Part B: Studies in History and Philosophy of Modern Physics}\ }\textbf {\bibinfo {volume} {39}},\ \bibinfo {pages} {782} (\bibinfo {year} {2008})}\BibitemShut {NoStop}%
\bibitem [{\citenamefont {Sutherland}(2017)}]{sutherland2017}%
  \BibitemOpen
  \bibfield  {author} {\bibinfo {author} {\bibfnamefont {R.~I.}\ \bibnamefont {Sutherland}},\ }\bibfield  {title} {\bibinfo {title} {Lagrangian description for particle interpretations of quantum mechanics: entangled many-particle case},\ }\href@noop {} {\bibfield  {journal} {\bibinfo  {journal} {Foundations of Physics}\ }\textbf {\bibinfo {volume} {47}},\ \bibinfo {pages} {174} (\bibinfo {year} {2017})}\BibitemShut {NoStop}%
\bibitem [{\citenamefont {Aharonov}\ and\ \citenamefont {Vaidman}(2008)}]{tsv}%
  \BibitemOpen
  \bibfield  {author} {\bibinfo {author} {\bibfnamefont {Y.}~\bibnamefont {Aharonov}}\ and\ \bibinfo {author} {\bibfnamefont {L.}~\bibnamefont {Vaidman}},\ }\bibfield  {title} {\bibinfo {title} {The two-state vector formalism: an updated review},\ }\href@noop {} {\bibfield  {journal} {\bibinfo  {journal} {Time in quantum mechanics}\ ,\ \bibinfo {pages} {399}} (\bibinfo {year} {2008})}\BibitemShut {NoStop}%
\bibitem [{\citenamefont {Aharonov}\ \emph {et~al.}(2015)\citenamefont {Aharonov}, \citenamefont {Cohen},\ and\ \citenamefont {Elitzur}}]{aharonov2015}%
  \BibitemOpen
  \bibfield  {author} {\bibinfo {author} {\bibfnamefont {Y.}~\bibnamefont {Aharonov}}, \bibinfo {author} {\bibfnamefont {E.}~\bibnamefont {Cohen}},\ and\ \bibinfo {author} {\bibfnamefont {A.~C.}\ \bibnamefont {Elitzur}},\ }\bibfield  {title} {\bibinfo {title} {Can a future choice affect a past measurement’s outcome?},\ }\href@noop {} {\bibfield  {journal} {\bibinfo  {journal} {Annals of Physics}\ }\textbf {\bibinfo {volume} {355}},\ \bibinfo {pages} {258} (\bibinfo {year} {2015})}\BibitemShut {NoStop}%
\bibitem [{\citenamefont {Wharton}(2014)}]{wharton2014}%
  \BibitemOpen
  \bibfield  {author} {\bibinfo {author} {\bibfnamefont {K.}~\bibnamefont {Wharton}},\ }\bibfield  {title} {\bibinfo {title} {Quantum states as ordinary information},\ }\href@noop {} {\bibfield  {journal} {\bibinfo  {journal} {Information}\ }\textbf {\bibinfo {volume} {5}},\ \bibinfo {pages} {190} (\bibinfo {year} {2014})}\BibitemShut {NoStop}%
\bibitem [{\citenamefont {Silberstein}\ \emph {et~al.}(2018)\citenamefont {Silberstein}, \citenamefont {Stuckey},\ and\ \citenamefont {McDevitt}}]{stuckey}%
  \BibitemOpen
  \bibfield  {author} {\bibinfo {author} {\bibfnamefont {M.}~\bibnamefont {Silberstein}}, \bibinfo {author} {\bibfnamefont {W.~M.}\ \bibnamefont {Stuckey}},\ and\ \bibinfo {author} {\bibfnamefont {T.}~\bibnamefont {McDevitt}},\ }\href@noop {} {\emph {\bibinfo {title} {Beyond the dynamical universe: unifying block universe physics and time as experienced}}}\ (\bibinfo  {publisher} {Oxford University Press},\ \bibinfo {year} {2018})\BibitemShut {NoStop}%
\bibitem [{\citenamefont {Adlam}(2022)}]{adlam2022}%
  \BibitemOpen
  \bibfield  {author} {\bibinfo {author} {\bibfnamefont {E.}~\bibnamefont {Adlam}},\ }\bibfield  {title} {\bibinfo {title} {Laws of nature as constraints},\ }\href@noop {} {\bibfield  {journal} {\bibinfo  {journal} {Foundations of Physics}\ }\textbf {\bibinfo {volume} {52}},\ \bibinfo {pages} {28} (\bibinfo {year} {2022})}\BibitemShut {NoStop}%
\bibitem [{\citenamefont {Chen}\ and\ \citenamefont {Goldstein}(2022)}]{chen2022}%
  \BibitemOpen
  \bibfield  {author} {\bibinfo {author} {\bibfnamefont {E.~K.}\ \bibnamefont {Chen}}\ and\ \bibinfo {author} {\bibfnamefont {S.}~\bibnamefont {Goldstein}},\ }\bibfield  {title} {\bibinfo {title} {Governing without a fundamental direction of time: Minimal primitivism about laws of nature},\ }in\ \href@noop {} {\emph {\bibinfo {booktitle} {Rethinking the Concept of Law of Nature: Natural Order in the Light of Contemporary Science}}}\ (\bibinfo  {publisher} {Springer},\ \bibinfo {year} {2022})\ pp.\ \bibinfo {pages} {21--64}\BibitemShut {NoStop}%
\bibitem [{\citenamefont {Wharton}(2015)}]{uinac}%
  \BibitemOpen
  \bibfield  {author} {\bibinfo {author} {\bibfnamefont {K.}~\bibnamefont {Wharton}},\ }\bibfield  {title} {\bibinfo {title} {The universe is not a computer},\ }in\ \href@noop {} {\emph {\bibinfo {booktitle} {Questioning the Foundations of Physics: Which of Our Fundamental Assumptions Are Wrong?}}}\ (\bibinfo  {publisher} {Springer},\ \bibinfo {year} {2015})\ pp.\ \bibinfo {pages} {177--189}\BibitemShut {NoStop}%
\bibitem [{\citenamefont {Braverman}\ and\ \citenamefont {Simon}(2013)}]{braverman}%
  \BibitemOpen
  \bibfield  {author} {\bibinfo {author} {\bibfnamefont {B.}~\bibnamefont {Braverman}}\ and\ \bibinfo {author} {\bibfnamefont {C.}~\bibnamefont {Simon}},\ }\bibfield  {title} {\bibinfo {title} {Proposal to observe the nonlocality of bohmian trajectories with entangled photons},\ }\href@noop {} {\bibfield  {journal} {\bibinfo  {journal} {Physical Review Letters}\ }\textbf {\bibinfo {volume} {110}},\ \bibinfo {pages} {060406} (\bibinfo {year} {2013})}\BibitemShut {NoStop}%
\bibitem [{\citenamefont {Norsen}(2017)}]{norsen}%
  \BibitemOpen
  \bibfield  {author} {\bibinfo {author} {\bibfnamefont {T.}~\bibnamefont {Norsen}},\ }\href@noop {} {\emph {\bibinfo {title} {Foundations of quantum mechanics}}}\ (\bibinfo  {publisher} {Springer},\ \bibinfo {year} {2017})\BibitemShut {NoStop}%
\bibitem [{\citenamefont {Catani}\ \emph {et~al.}(2023)\citenamefont {Catani}, \citenamefont {Leifer}, \citenamefont {Schmid},\ and\ \citenamefont {Spekkens}}]{leifer}%
  \BibitemOpen
  \bibfield  {author} {\bibinfo {author} {\bibfnamefont {L.}~\bibnamefont {Catani}}, \bibinfo {author} {\bibfnamefont {M.}~\bibnamefont {Leifer}}, \bibinfo {author} {\bibfnamefont {D.}~\bibnamefont {Schmid}},\ and\ \bibinfo {author} {\bibfnamefont {R.~W.}\ \bibnamefont {Spekkens}},\ }\bibfield  {title} {\bibinfo {title} {Why interference phenomena do not capture the essence of quantum theory},\ }\href@noop {} {\bibfield  {journal} {\bibinfo  {journal} {Quantum}\ }\textbf {\bibinfo {volume} {7}},\ \bibinfo {pages} {1119} (\bibinfo {year} {2023})}\BibitemShut {NoStop}%
\bibitem [{\citenamefont {Wharton}(2018)}]{wharton2018}%
  \BibitemOpen
  \bibfield  {author} {\bibinfo {author} {\bibfnamefont {K.}~\bibnamefont {Wharton}},\ }\bibfield  {title} {\bibinfo {title} {A new class of retrocausal models},\ }\href@noop {} {\bibfield  {journal} {\bibinfo  {journal} {Entropy}\ }\textbf {\bibinfo {volume} {20}},\ \bibinfo {pages} {410} (\bibinfo {year} {2018})}\BibitemShut {NoStop}%
\bibitem [{\citenamefont {Wharton}\ and\ \citenamefont {Koch}(2015)}]{wharton2015}%
  \BibitemOpen
  \bibfield  {author} {\bibinfo {author} {\bibfnamefont {K.}~\bibnamefont {Wharton}}\ and\ \bibinfo {author} {\bibfnamefont {D.}~\bibnamefont {Koch}},\ }\bibfield  {title} {\bibinfo {title} {Unit quaternions and the bloch sphere},\ }\href@noop {} {\bibfield  {journal} {\bibinfo  {journal} {Journal of Physics A: Mathematical and Theoretical}\ }\textbf {\bibinfo {volume} {48}},\ \bibinfo {pages} {235302} (\bibinfo {year} {2015})}\BibitemShut {NoStop}%
\bibitem [{\citenamefont {Einstein}\ \emph {et~al.}(1935)\citenamefont {Einstein}, \citenamefont {Podolsky},\ and\ \citenamefont {Rosen}}]{einstein1935}%
  \BibitemOpen
  \bibfield  {author} {\bibinfo {author} {\bibfnamefont {A.}~\bibnamefont {Einstein}}, \bibinfo {author} {\bibfnamefont {B.}~\bibnamefont {Podolsky}},\ and\ \bibinfo {author} {\bibfnamefont {N.}~\bibnamefont {Rosen}},\ }\bibfield  {title} {\bibinfo {title} {Can quantum-mechanical description of physical reality be considered complete?},\ }\href@noop {} {\bibfield  {journal} {\bibinfo  {journal} {Physical review}\ }\textbf {\bibinfo {volume} {47}},\ \bibinfo {pages} {777} (\bibinfo {year} {1935})}\BibitemShut {NoStop}%
\bibitem [{\citenamefont {Bell}(2004)}]{bell2004}%
  \BibitemOpen
  \bibfield  {author} {\bibinfo {author} {\bibfnamefont {J.~S.}\ \bibnamefont {Bell}},\ }\href@noop {} {\emph {\bibinfo {title} {Speakable and unspeakable in quantum mechanics: Collected papers on quantum philosophy}}}\ (\bibinfo  {publisher} {Cambridge university press},\ \bibinfo {year} {2004})\BibitemShut {NoStop}%
\bibitem [{\citenamefont {Neder}\ and\ \citenamefont {Argaman}(2024)}]{neder2024}%
  \BibitemOpen
  \bibfield  {author} {\bibinfo {author} {\bibfnamefont {I.}~\bibnamefont {Neder}}\ and\ \bibinfo {author} {\bibfnamefont {N.}~\bibnamefont {Argaman}},\ }\bibfield  {title} {\bibinfo {title} {Future-input-dependent model for greenberger-horne-zeilinger correlations},\ }\href@noop {} {\bibfield  {journal} {\bibinfo  {journal} {Physical Review A}\ }\textbf {\bibinfo {volume} {110}},\ \bibinfo {pages} {032209} (\bibinfo {year} {2024})}\BibitemShut {NoStop}%
\bibitem [{\citenamefont {Wharton}\ and\ \citenamefont {Adlam}(2023)}]{whartonadlam}%
  \BibitemOpen
  \bibfield  {author} {\bibinfo {author} {\bibfnamefont {K.}~\bibnamefont {Wharton}}\ and\ \bibinfo {author} {\bibfnamefont {E.}~\bibnamefont {Adlam}},\ }\bibfield  {title} {\bibinfo {title} {Entangled photon anti-correlations are evident from classical electromagnetism},\ }\href@noop {} {\bibfield  {journal} {\bibinfo  {journal} {Symmetry}\ }\textbf {\bibinfo {volume} {15}},\ \bibinfo {pages} {1539} (\bibinfo {year} {2023})}\BibitemShut {NoStop}%
\end{thebibliography}%

\onecolumngrid
\appendix

\section{}

Here we demonstrate that Eq. (\ref{eq:eoms}) follows from (\ref{eq:weak}) and (\ref{eq:Uex}).  We need only prove that
\begin{equation}
\label{eq:prove}
    \frac{d^2 w_a}{d \tau ^2} = \frac{\pi^2}{2}(w_b - w_a),
\end{equation}
where $w_a$ and $w_b$ are any arbitrary component of the full vectors $\bm{w}_a$ and $\bm{w}_b$.  Also, the second equation in (\ref{eq:eoms}) follows from interchanging the two qubits.  

Consider the limit definition of the second derivative:
\begin{equation}
\label{eq:p1}
    \frac{d^2 w_a}{d \tau ^2} = \lim_{\Delta \tau \to 0} \frac{w_a(\tau + \Delta \tau) - 2w_a(\tau) + w_a(\tau - \Delta \tau)}{\Delta \tau^2}
\end{equation}
The exchange interaction for a generic $SWAP^\alpha$ gate runs to a time $\tau=\alpha$, so these weak value terms can be found from:
\begin{eqnarray}
     w_a(\tau + \Delta \tau) &=& \frac{\langle f|U_{ex}(\alpha - \tau)U_{ex}^\dagger(\Delta \tau) (\sigma \otimes I) U_{ex}(\Delta \tau)U_{ex}(\tau) | i \rangle}{\langle f|U_{ex}(\alpha)|i \rangle}\nonumber \\
     w_a(\tau) &=&  \frac{\langle f|U_{ex}(\alpha - \tau)(\sigma \otimes I) U_{ex}(\tau)| i \rangle}{\langle f|U_{ex}(\alpha)|i \rangle} \nonumber\\
     w_a(\tau - \Delta \tau) &=&  \frac{\langle f|U_{ex}(\alpha - \tau) U_{ex}(\Delta \tau) (\sigma \otimes I) U_{ex}^\dagger(\Delta \tau)U_{ex}(\tau)| i \rangle}{\langle f|U_{ex}(\alpha)|i \rangle}  \nonumber
\end{eqnarray}
where $U_{ex}^\dagger(\Delta \tau) = U_{ex}(-\Delta \tau) = U_{ex}^{-1}(\Delta \tau)$, and $\sigma$ is any of the Pauli matrices $\sigma_x $, $\sigma_y $, $\sigma_z $ (corresponding to the component of interest of $\bm{w}_a$ and $\bm{w}_b$).
Then, setting $\langle f'| = \langle f|U_{ex}(\alpha - \tau)$ and $U_{ex} (\tau)|i\rangle = |i'\rangle$, Eq. (\ref{eq:p1}) becomes:
\begin{equation}
\label{eq:p2}
     \frac{d^2 w_a}{d \tau ^2} = \lim_{\Delta \tau \to 0} \frac{\langle f'|U_{ex}^\dagger(\Delta \tau) (\sigma \otimes I) U_{ex}(\Delta \tau) - 2(\sigma \otimes I) + U_{ex}(\Delta \tau) (\sigma \otimes I) U_{ex}^\dagger(\Delta \tau) |i'\rangle}{{\Delta \tau^2}\langle f|U_{ex}(\alpha)|i \rangle}.
\end{equation}

Expanding $U_{ex}(\Delta \tau)$ to second order in $\Delta \tau$, it happens that
\begin{align*}
    U_{ex}(\Delta \tau) \approx  I + \left( i \frac{\pi \Delta \tau}{2} - \frac{\pi^2 \Delta \tau^2}{4} \right)  \Big ( \frac{I - \sigma_x \otimes \sigma_x -\sigma_y \otimes \sigma_y - \sigma_z \otimes \sigma_z}{2} \Big ).
\end{align*}
Defining $M=(I-\bm{\sigma}\otimes\bm{\sigma})/2$, and a complex constant $a$, this can be expressed as $U_{ex}=I+aM$, dropping terms of order $(\Delta\tau)^3$ and higher.

With these definitions, (\ref{eq:p2}) then becomes:
\begin{equation}
\frac{d^2 w_a}{d \tau ^2} = \frac{\langle f'| 2 aa^*  M(\sigma \otimes I) M + (a + a^*) ( M (\sigma \otimes I) + (\sigma \otimes I) M )|i'\rangle } {{\Delta \tau^2}\langle f'|i' \rangle}.
\end{equation}
This expression greatly simplifies, because of the easily-checked relationships
\begin{eqnarray}
    M(\sigma \otimes I) M &=& 0\\
    a + a^* &=& -\frac{\pi^2 \Delta \tau^2}{2}\\
    M (\sigma \otimes I) + (\sigma \otimes I) M &=& (\sigma \otimes I) - (I \otimes \sigma).
\end{eqnarray}
Again, this final equation is correct for any given component $\sigma$ of the Pauli vector $\bm{\sigma}$.

Finally, using the fact that $w_b(\tau) =\bra{f'} I \otimes \sigma\ket{i'}/\braket{f'|i'}$, we find the desired equation
\begin{align*}
    \frac{d^2 w_a}{d \tau ^2} =  \frac{\pi^2 \Delta \tau^2 \langle f'| I \otimes \sigma - \sigma \otimes I |i'\rangle}{2\Delta \tau^2 \langle f'|i' \rangle}  = \frac{\pi^2}{2} (w_b - w_a)
\end{align*}

\newpage

\section{}

Inside the $\sqrt{SWAP}$ gate of Figure 4, the closed-form expressions for the complex weak value vectors $\bm{w_a}$ and $\bm{w_b}$ are given below, depending on the outcome state $\ket{f}$.  Here $\tau=0$ is the input (left side of the gate) and $\tau=0.5$ corresponds to the output (right side of the gate).  The continuous evolution inside the gate can be found by varying $\tau$.

For $\ket{f}=\ket{00}$:
\begin{eqnarray}
    \bm{w}_a &=& \frac{1}{2} \left(1+i+(1-i)e^{i\pi\tau}\;,\;1-i-(1+i) e^{i\pi\tau}\;,\;2 \right) \\
    \bm{w}_b &=& \frac{1}{2} \left(1+i-(1-i)e^{i\pi\tau}\;,\;1-i+(1+i) e^{i\pi\tau}\;,\;2 \right)
\end{eqnarray}

For $\ket{f}=\ket{10}$:
\begin{eqnarray}
    \bm{w}_a &=& \frac{1}{2} \left(1+e^{-i\pi\tau}\;,\;1- e^{-i\pi\tau}\;,\; -2 \sin(\pi\tau) \right) \\
    \bm{w}_b &=& \frac{1}{2} \left(1-e^{-i\pi\tau}\;,\;1+ e^{-i\pi\tau}\;,\; 2 \sin(\pi\tau)  \right)
\end{eqnarray}

And for $\ket{f}=\ket{11}$:
\begin{eqnarray}
    \bm{w}_a &=& \frac{1}{2} \left(1-i+(1+i)e^{i\pi\tau}\;,\;1+i-(1-i) e^{i\pi\tau}\;,\;-2 \right) \\
    \bm{w}_b &=& \frac{1}{2} \left(1-i-(1+i)e^{i\pi\tau}\;,\;1+i+(1-i) e^{i\pi\tau}\;,\;-2 \right)
\end{eqnarray}
Notice in all cases, one finds $\ddot{\bm{w}}_a=(\pi^2/2)(\bm{w_b}-\bm{w_a})$ and $\ddot{\bm{w}}_b=(\pi^2/2)(\bm{w_a}-\bm{w_b})$.  Also, for all cases, $\dot{\bm{w}}_a = -\dot{\bm{w}}_b$.  But only for the $\ket{f}=\ket{00}$ and $\ket{f}=\ket{11}$ case does $\dot{\bm{w}}_a = (\pi/2)\bm{w_b}\times\bm{w_a}.$

\end{document}